\documentclass[iop]{emurateapj}

\usepackage{bm}

\shorttitle{The Cosmogrid Simulation}
\shortauthors{Ishiyama et al.}

\begin{document}

\title{The Cosmogrid Simulation: Statistical Properties of Small Dark Matter Halos}

\author{ 
TOMOAKI \textsc{Ishiyama}\altaffilmark{1}, 
STEVEN \textsc{Rieder}\altaffilmark{2,3}, 
JUNICHIRO \textsc{Makino}\altaffilmark{4,5}, 
SIMON \textsc{Portegies Zwart}\altaffilmark{2}, 
DEREK \textsc{Groen}\altaffilmark{6}, 
KEIGO \textsc{Nitadori}\altaffilmark{5}, 
CEES \textsc{de Laat}\altaffilmark{3},
STEPHEN \textsc{McMillan}\altaffilmark{7}, 
KEI \textsc{Hiraki}\altaffilmark{8}, and 
STEFAN \textsc{Harfst}\altaffilmark{9} 
} 

\altaffiltext{1}{Center for Computational Science, University of Tsukuba, 1-1-1, Tennodai, Tsukuba, Ibaraki, 305-8577, Japan ; ishiyama@ccs.tsukuba.ac.jp}
\altaffiltext{2}{Sterrewacht Leiden, Leiden University, P.O. Box 9513, 2300 RA Leiden, The Netherlands} 
\altaffiltext{3}{Section System and Network Engineering, University of Amsterdam, Amsterdam, The Netherlands}
\altaffiltext{4}{Graduate School of Science and Engineering, Tokyo Institute of Technology, Japan}
\altaffiltext{5}{RIKEN Advanced Institute for Computational Science, Japan}
\altaffiltext{6}{Centre for Computational Science, Department of Chemistry, University College London, 20 Gordon Street, London, WC1H 0AJ, UK}
\altaffiltext{7}{Department of Physics, Drexel University, Philadelphia, PA 19104, USA}
\altaffiltext{8}{Department of Creative Informatics, Graduate School of Information Science and Technology the University of Tokyo, Japan}
\altaffiltext{9}{Center for Astronomy and Astrophysics, Technical University Berlin, Hardenbergstr. 36, D-10623 Berlin, Germany}

\begin{abstract}
We present the results of the ``Cosmogrid'' cosmological
$N$-body simulation suites based on the concordance LCDM model.  The
Cosmogrid simulation was performed in a 30Mpc box with $2048^3$
particles.
The mass of each particle is $1.28 \times 10^5 M_{\odot}$, 
which is sufficient to resolve ultra-faint dwarfs.
We found that the halo mass function shows good agreement with the
Sheth \& Tormen fitting function down to $\sim 10^7 M_{\odot}$.
We have analyzed the spherically averaged density profiles of the three most
massive halos which are of galaxy group size and contain at least 170 million
particles.  
The slopes of these density profiles
become shallower than $-1$ at the inner most radius.
We also find a clear correlation of halo concentration with mass. 
The mass dependence of the concentration parameter 
cannot be expressed by a single power law, however 
a simple model based on the Press$-$Schechter theory 
proposed by Navarro et al. 
gives reasonable agreement with this dependence.
The spin parameter does not show a correlation with the halo mass.  
The probability distribution functions for both concentration and spin are well
fitted by the log-normal distribution for halos with the masses larger
than $\sim 10^8 M_{\odot}$.
The subhalo abundance depends on the halo mass. 
Galaxy-sized halos have 50\% more subhalos than 
$\sim 10^{11} M_{\odot}$ halos have.
\end{abstract}


\section{Introduction}
According to the present standard LCDM model, 
the universe is thought to be composed primarily of 
cold dark matter (CDM) and dark energy \citep{White1978, Peacock1999}.  
Structure formation of the universe proceeds hierarchically in this model.  
Smaller-scale
structures collapse first, and then merge into larger-scale structures. 

There is serious discrepancy between the distribution of subhalos
in galaxy-sized halos obtained by numerical simulations and the
observed number of dwarf galaxies in the Local Group
\citep{Klypin1999, Moore1999a}.
This ``missing dwarf problem'' is still considered to be one of the
most serious problems in the CDM paradigm
\citep[e.g.,][]{Kroupa2010}. 
In order to understand the
origin of this discrepancy, it is necessary to perform high-resolution
cosmological $N$-body simulations 
and obtain unbiased sample of galaxy-sized halos with
resolution high enough to obtain reliable statistics of subhalos 
since the subhalo abundance shows large halo-to-halo variations 
\citep{Ishiyama2009}.

Cosmological $N$-body simulations have been widely used to study the
nonlinear structure formation of the universe and 
have been an important tool for a better understanding of our universe. 
In order to study the spatial correlation of galaxies, 
the first cosmological $N$-body simulations were performed in the 1970s
using approximately 1000 particles
\citep[e.g.,][]{Miyoshi1975, Fall1978, Aarseth1979, Efstathiou1979}.
Since then, the development of better simulation algorithms and improvements 
in the performance of computers  allow us to use much larger numbers of particles
and have drastically increased the resolution of cosmological simulations.

Today, it is not uncommon that the number of particles exceeds $10^9$ 
in high-resolution simulations.  
In these works, the size of the simulation volumes is typically $[O(\rm Gpc)]^3$ 
and populations of galaxy clusters, gravitational lensing, 
and the baryon acoustic oscillation are studied
\citep[e.g.,][]{Evrard2002, Wambsganss2004, Teyssier2009,
Kim2009, Crocce2010}.
The simulation results are also used to construct 
mock halo catalogs for next generation large
volume surveys.  
Others use simulations of $[O(\rm 100Mpc)]^3$ volumes
to study the internal properties of galaxy-sized dark matter halos,
their formation, evolution, and statistical properties
\citep[e.g.,][]{Springel2005, Klypin2011, White2010}.

Using the results of high-resolution simulations of small-scale structures, 
we can study the fine structures of galactic halos,
the distribution of subhalos, their structures, 
and their dependence on the nature of dark matter. 
This information has a strong impact on the indirect search
for dark matter
since gamma-ray flux by self-annihilation is proportional 
to local density if we consider neutralino as the candidate of dark matter. 
Thus, we can restrict the nature of dark matter 
using the results of  high-resolution simulations of small-scale structures 
and indirect searches of dark matter.
In addition, galaxies are considered to form in dark matter halos with a mass
larger than a critical value \citep{Strigari2008, Li2009, Maccio2009, Okamoto2009}.  
The structure of smallest halos which can host
galaxies is important for the understanding of the galaxy formation processes.

The simulation of smaller-scale structures of dark matter halos is
not a trivial task since a very wide dynamic range of space, mass, and
time must be covered. 
In particular, the number of time steps of such simulations 
is significantly larger than that of larger-scale simulations
since the dynamical time-scale is proportional to $1.0/\sqrt{G\bar{\rho}}$, 
where $\bar{\rho}$ is the local density.
Structures of smaller scales form earlier, and thus have higher densities, 
therefore, simulations of smaller scales are computationally more expensive.

Recently, simulations with galactic halos of very high-resolution 
have been performed \citep{Diemand2008, Springel2008, Stadel2009}.
These works used the re-simulation method, where one selects one or a
few halos at $z=0$ from a simulation which covers a large volume 
(typically a cube of size O(100Mpc)) with a relatively
low-resolution. 
The corresponding regions of these halos are then
identified in the initial particle distribution, 
and the particles in these regions are replaced by a larger number of
smaller particles. 
After this is done, the entire volume is simulated to $z=0$ again.

With this re-simulation method, 
we can resolve the structures of selected halos with extremely high
resolution \citep{Diemand2008, Springel2008, Stadel2009}.  
However, this method cannot be used for the study of halo-to-halo variations.
Different halos are born in different environments and 
grow differently. 
The difference in the environment and growth history 
must be the cause of halo-to-halo variations.
Therefore, in order to study variations, 
we need a bias-free set of a large number of halos.
Clearly one cannot obtain a large number of halos with 
re-simulation method in practical time.

In principle, one can improve the statistics by increasing 
the number of halos selected for re-simulations.
In order to avoid the selection bias, 
we need to apply random, bias-free selection, 
and the most reliable bias-free selection is 
to select all halos, in other words, 
to simulate the entire simulation box with uniformly high mass resolution.
\citet{Ishiyama2009} performed the first bias-free high
resolution simulation of small-scale structures.
They analyzed the statistics of the subhalo abundance 
using the complete set of halos in the simulation box. 
The number of particles was $1600^3$ in a 46.5Mpc cubic box
and the mass of a particle was $10^6 M_{\odot}$.  
The subhalo abundance showed large halo-to-halo variations (see also
\citep{Ishiyama2008, Boylan2010}).  The concentration
parameter and the radius at the moment of the maximum expansion showed
fairly a tight correlation with the subhalo abundance.  Halos formed
earlier have a smaller number of subhalos at present.  This correlation
suggests that the difference in the formation history is the origin
of the variation of the subhalo abundance 
(see also \citet{Gao2004, Bosch2005, Zentner2005}).

The Millennium-II simulation \citep{Boylan2009} used a 137Mpc cubic box
and the particle mass of $\sim 9.45 \times 10^6 M_{\odot}$.  
Its result is suitable for the analysis of 
the statistics of galaxy-sized dark matter halos,
because the number of halos is larger than that 
of \citet{Ishiyama2009}.
However, due to the lack of the mass resolution,
it cannot be used to 
study the statistics of dwarf-galaxy-sized 
halos and the statistics of subhalos with the size
larger than faint dwarf galaxy.

In this paper, we describe the first result of our Cosmogrid simulation.
We simulated the evolution of halos in a 30Mpc cubic box using $2048^3$ particles.
The mass of one particle is $1.28 \times 10^5 M_{\odot}$.
The resolution reaches down to ultra-faint dwarf-galaxy-sized halos
($\sim 10^7 M_{\odot}$)
and is more than eight times better 
than that of our previous simulation \citep{Ishiyama2009}.  
We focus on the halo mass function with the mass down to
$10^7 M_{\odot}$, the structures of most massive halos, 
and statistics of the internal
properties of dwarf-galaxy-sized halos.  We describe our initial
conditions and numerical settings in Section \ref{sec:method}, 
and results in Section \ref{sec:result}.  We discuss and summarize our
results in Section \ref{sec:discussion}.

\section{Initial Conditions and Numerical Method}\label{sec:method}

The cosmological parameters adopted are based on the concordance LCDM
cosmological model ($\Omega_0=0.3$, $\lambda_0=0.7$, $h=0.7$,
$\sigma_8=0.8$, $n=1.0$).  These values are the same as 
those used in our previous simulation \citep{Ishiyama2009}.  
We used a periodic cube of the
comoving size of 30Mpc.  The number of particles for the largest run
is $2048^3$ which corresponds to a mass resolution of $1.28 \times 10^5 M_{\odot}$.
To generate the initial particle distributions, we used the
MPGRAFIC package \citep{Prunet2009}, which is a parallelized variation
of the GRAFIC package \citep{Bertschinger2001}.  The initial red
was 65.

In order to investigate the effect of the mass and spatial resolution, 
we performed two simulations with lower resolution. 
We generated the initial conditions for these low-resolution runs (CG1024 and CG512)
by replacing 8 or 64 particles in the high-resolution initial condition (CG2048)
with a single particle 8 or 64 times more massive.
We did not use any smoothing filter for density and velocity spaces. 
The massive particles were picked up at regular intervals before 
performing the Zel'dovich approximation.
This procedure introduces some aliasing noise in the high frequency limit 
of CG1024 and CG512 runs. 
The corresponding halo contains less than a few hundred particles.
However, here we use CG1024 and CG512 runs for only convergence studies, 
and analyze halos with the particles larger than $\sim1000$. 
Thus, the effect of the aliasing noise should be negligible.
In Table \ref{tab1}, we summarize parameters used in our simulations.

\begin{table*}[t]
\centering
\caption{Run Parameters\label{tab1}. 
Here, $N$, $L$, $\varepsilon$, and $m$ are the total number of particles, 
the box length, the softening length, the mass resolution.
}
\begin{tabular}{lcccc}
\hline\hline
Name  & $N$ & $L(\rm Mpc)$ & $\varepsilon(\rm pc)$ & $m(\rm M_{\odot})$\\
\hline
CG2048    & $2048^3$ & 30.0 & 175 & $1.28 \times 10^5$\\
CG1024    & $1024^3$ & 30.0 & 350 & $1.03 \times 10^6$\\
CG512     & $512^3$  & 30.0 & 700 & $8.21 \times 10^6$\\
IFM2009 \citep{Ishiyama2009} & $1600^3$  & 46.5 & 700 & $1.00 \times 10^6$\\
\hline 
\end{tabular}
\end{table*}

\begin{table*}[t]
\centering
\caption
{ Global Parameters of Three Most Massive Group Zized Halos at $z=0$.
Here, $M$, $N$, $R_{\rm vir}$, $R_{\rm vmax}$, and $V_{\rm max}$ are
the mass, the number of particles,
the virial radius
in which the spherical overdensity is 101 times the critical value, 
the radius where the rotation velocity is maximum, 
and the maximum rotation velocity, respectively.
}
\label{tab2}
\begin{tabular}{lcccccc}
\hline \hline
Name & Run & $M (10^{13}M_{\odot})$ & $N$ & $R_{\rm vir} ({\rm kpc})$ & $R_{\rm vmax} ({\rm kpc})$ & $V_{\rm max} ({\rm kms^{-1}})$  \\
\hline
GP1 & CG2048 & 5.24 & 408499843 & 969 & 200 & 596 \\
    & CG1024 & 5.19 & 50632942 & 966 & 186 & 589 \\
    & CG512  & 5.22 & 6361253 & 968 & 184 & 596 \\
\hline
GP2 & CG2048 & 3.58 & 279382586 & 854 & 305 & 476 \\
    & CG1024 & 3.57 & 34836692 & 853 & 279 & 472 \\
    & CG512  & 3.57 & 4347651 & 852 & 294 & 475 \\
\hline
GP3 & CG2048 & 2.25 & 175752770 & 731 & 178 & 434 \\
    & CG1024 & 2.26 & 22072073 & 732 & 187 & 431 \\
    & CG512  & 2.25 & 2746874 & 731 & 192 & 434 \\
\hline
\end{tabular}
\end{table*}

We used a leapfrog integrator with shared and adaptive time steps.  The
step size was determined as $\min(2.0 \sqrt{\varepsilon/|\bm{a}_i|},2.0
\varepsilon/|\bm{v}_i|)$ (minimum of these two values for all
particles).  
All particles have the same timesteps.
The gravitational plummer softening length $\varepsilon$
was $\rm 175pc$ at $z=0$.  The softening was constant in comoving coordinates from
$z=65$ (initial condition) to $z=10$.  
From $z=10$ to $z=0$, it was constant in physical coordinates.
This procedure is similar to that used in \citet{Kawai2004}.

For the largest simulation, we used four supercomputers.
Three of them are Cray XT4 machines at the Center for
Computational Astrophysics of National Astronomical Observatory of
Japan, the Edinburgh Parallel Computing Center in Edinburgh (United
Kingdom) and IT Center for Science in Espoo (Finland).  
The fourth machine is an IBM pSeries 575 
at SARA in Amsterdam (the Netherlands).  
Part of the calculation was done in a ``grid'' computing environment, 
in which we used more than one machine simultaneously for one run 
\citep{Zwart2010}. 

For the time integration we used the GreeM code \citep{Ishiyama2009b} for
single supercomputer runs and the SUSHI code \citep{Groen2011}
for multi-supercomputer runs.  The GreeM code is a massively parallel
TreePM code based on the parallel TreePM code of \citet{Yoshikawa2005}
for large cosmological $N$-body simulations.  The long range forces are
calculated by the PM method \citep{Hockney1981}, and the short range forces
are calculated by the Barnes-Hut-Tree method \citep{Barnes1986}.  
\citet{Yoshikawa2005} used a 1-D slab decomposition, but in GreeM 
we use a 3-D multi-section decomposition \citep{Makino2004} to improve its scalability.
In addition, the decomposition is based on CPU time measurement, 
so that near ideal load balance is archived.
The SUSHI code is an extension of the GreeM code which can run 
concurrently on multiple supercomputers.
It uses the MPWide communication library \citep{Groen2010}
to facilitate message passing between distributed supercomputers.
We used $512^3$ PM grid points for PM calculations, the
opening angle for the tree method was 0.3 from initial to $z=10$, and
0.5 from $z=10$ to $z=0$.  

The calculation time was $\sim$180s per step with 1024 cpu cores
for the largest run on the Cray XT4 in Japan and 
$\sim$140s per step with 2048 cpu cores
on the IBM pSeries 575 in the Netherlands. 
We spent about 3.5 million CPU hours to perform all the 
60,283 steps in our simulation.

We used the spherical overdensity method \citep{Lacey1994} to 
identify halos and calculated the halo virial radius $R_{\rm vir}$.
The virial radius of a halo is defined as the radius
in which the spherical overdensity is $\Delta(z)$ times the critical
 value.  
The overdensity $\Delta(z)$ is given by the analytic formula
\citep{Bryan1998},
\begin{eqnarray}
\Delta(z) = (18\pi^2+ 82x - 39x^2)/\Omega(z), \label{eq:overdensity}
\end{eqnarray}
where $x \equiv \Omega(z)-1$.  The mass of
a halo is defined as interior mass within the virial radius.

The mass of the most massive halo is $5.24 \times
10^{13} M_{\odot}$. It contains $4.08 \times 10^{8}$ particles.
Via Lactea I, II \citep{Diemand2007, Diemand2008}, 
and Aquarius simulations \citep{Springel2008}
used ${\sim 10^{8}}$, ${\sim 5 \times 10^{8}}$, and ${\sim 10^{9}}$ particles
for the largest halo.
Table \ref{tab2} shows the
properties of the three most massive halos in our simulation.  

The subhalo finder is the
  same as that described in \citet{Ishiyama2009}. Our method is
  based on the idea of finding all local potential minima.  Initially, all
  particles are candidates for the centers of halos.  
  We then search for the
  particle with the smallest (most negative) potential 
  and regard it as 
  the center of a halo.  We then exclude $n_{\rm min}$ neighbor
  particles of this particle from the list of remaining particles, and search
  the particle with the smallest potential from the list.
  At this time, we again search $n_{\rm min}$ neighbor
  particles from the list of originally selected particles, and if the
  potential of one neighbor is smaller, we do not add this particle
  to the list of halos.  However, we remove $n_{\rm min}$ neighbors no
  matter whether the particle is added to the list or not.  We repeat this
  procedure until there is no remaining particle.  We set $n_{\rm
    min}$ so that $n_{\rm min} \times m = 1.0\times 10^{7}M_{\odot}$, 
where $m$ is the mass of each particle.

\begin{figure*}
\centering 
\includegraphics[width=14cm]{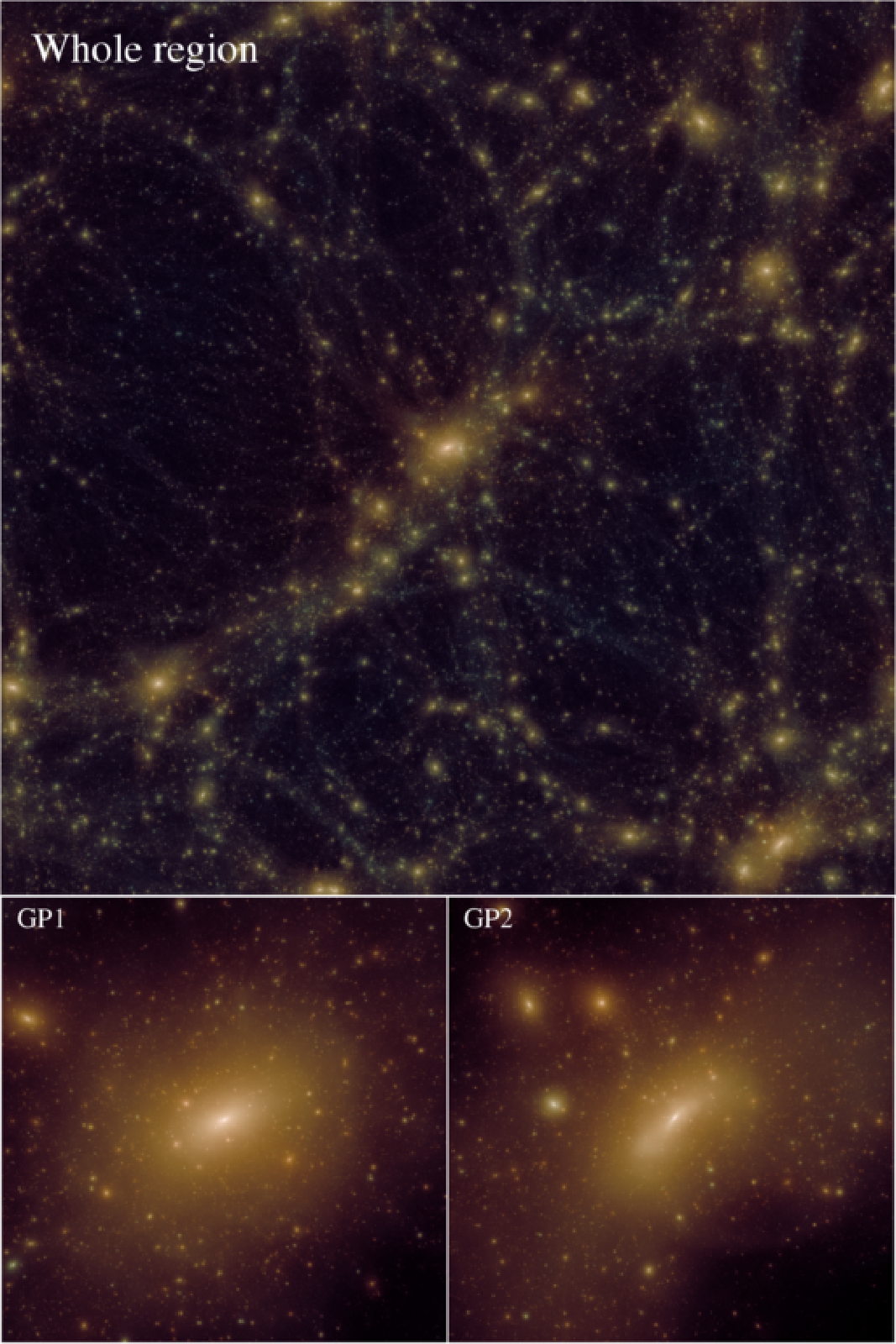} 
\caption{ 
Projected density of dark matter at $z=0$ in our largest simulation
($2048^3$ particles).  Top panel shows the whole region with the volume
of $\rm (30Mpc)^3$.  Bottom panels show the projected density of the two
most massive group sized halos.  These volumes are $\rm (2Mpc)^3$.  
}
\label{fig:snapshot1}
\end{figure*}

\begin{figure*}
\centering
\includegraphics[width=16cm]{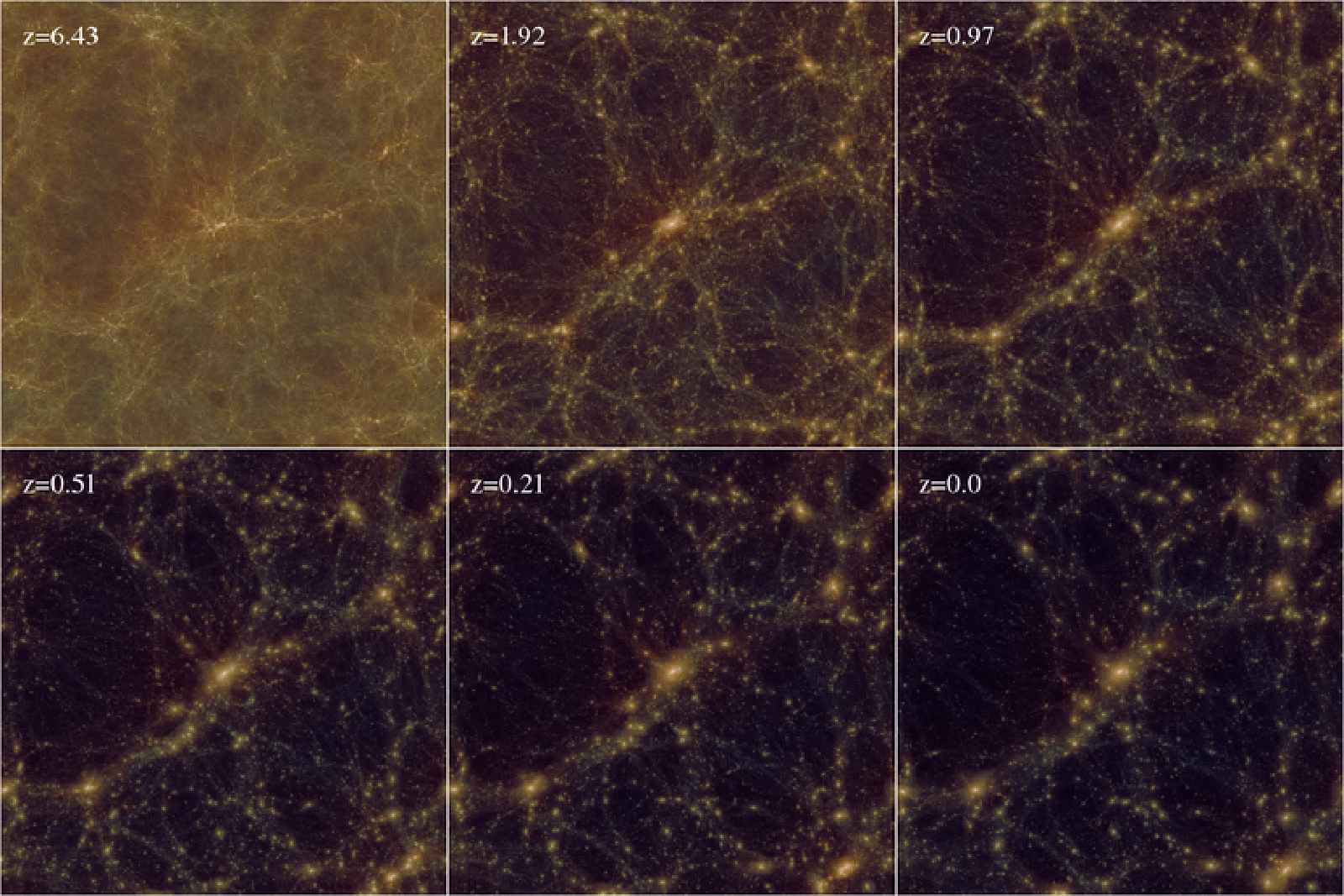}
\includegraphics[width=16cm]{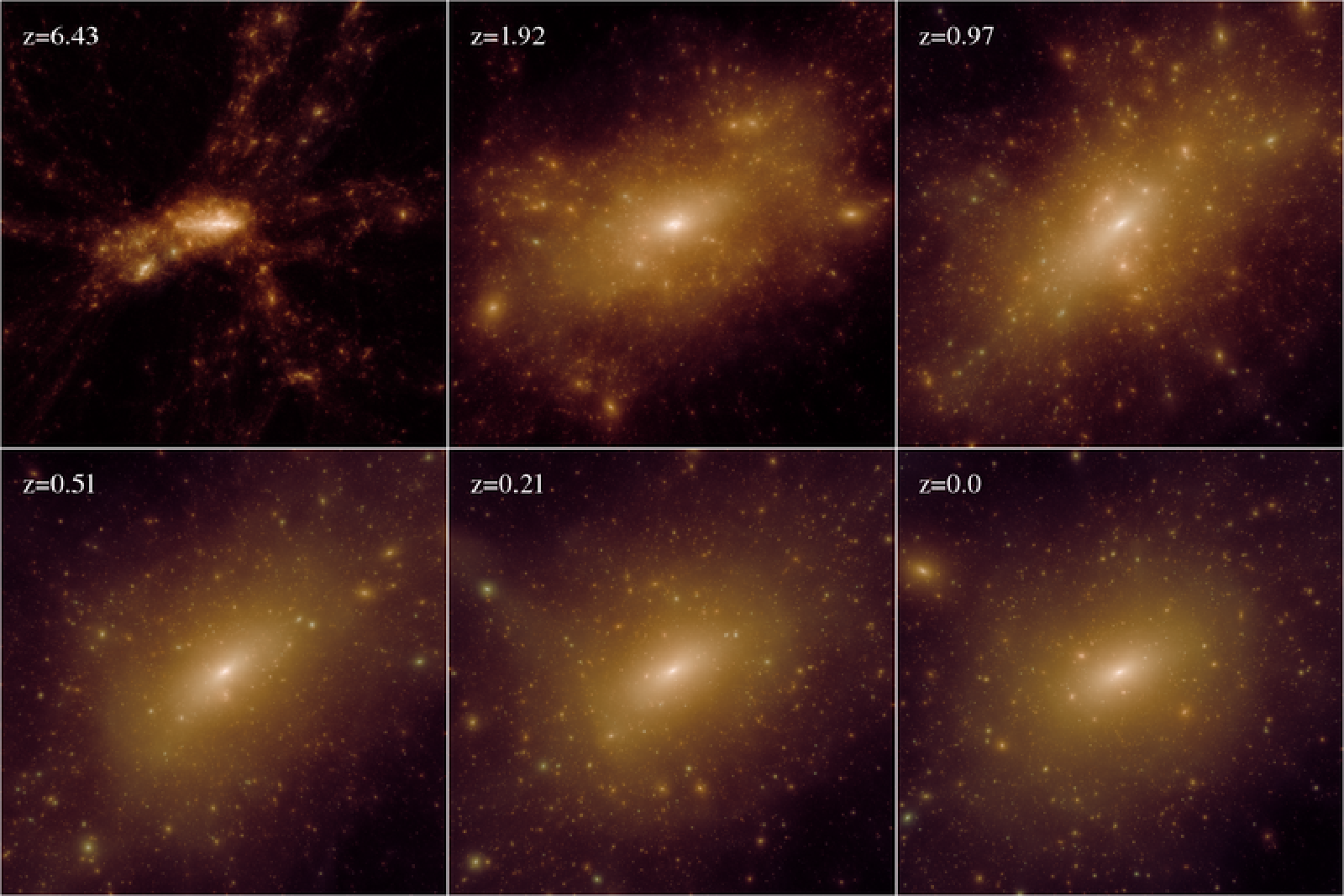}
\caption{
Evolution pictures of our largest simulation.
Top six panels show the evolution of the whole region. 
Bottom six panels show the evolution of the most massive halo.
}
\label{fig:snapshot2}
\end{figure*}

\begin{figure*}
\centering
\includegraphics[width=16cm]{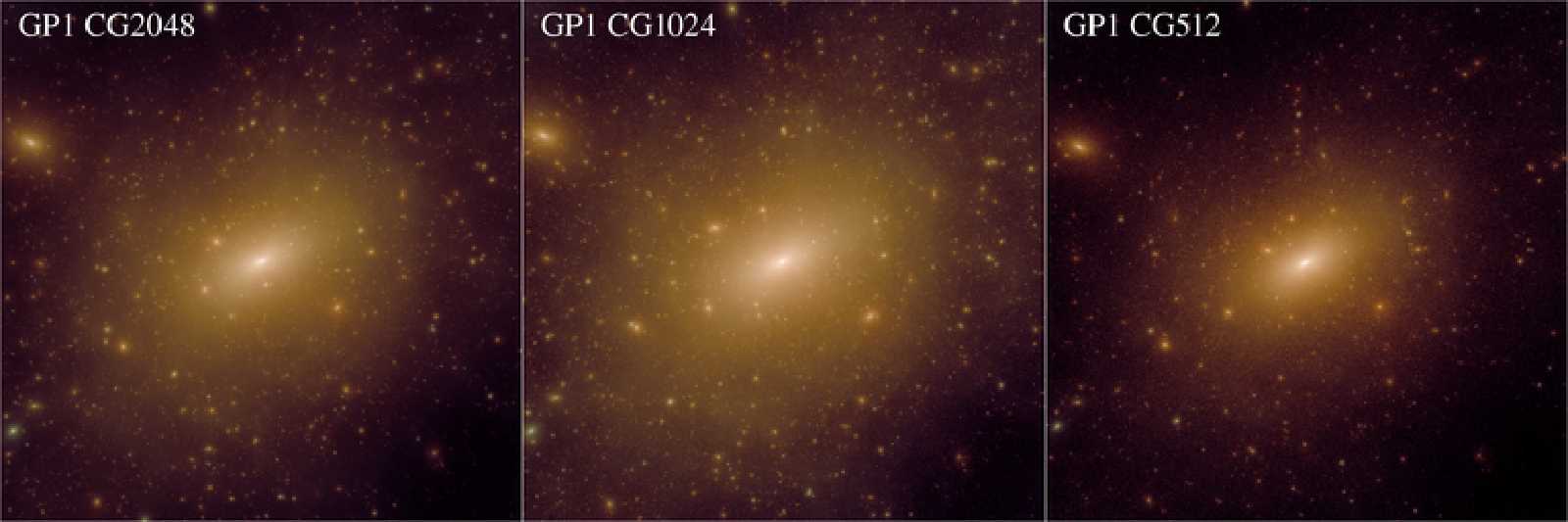}
\includegraphics[width=16cm]{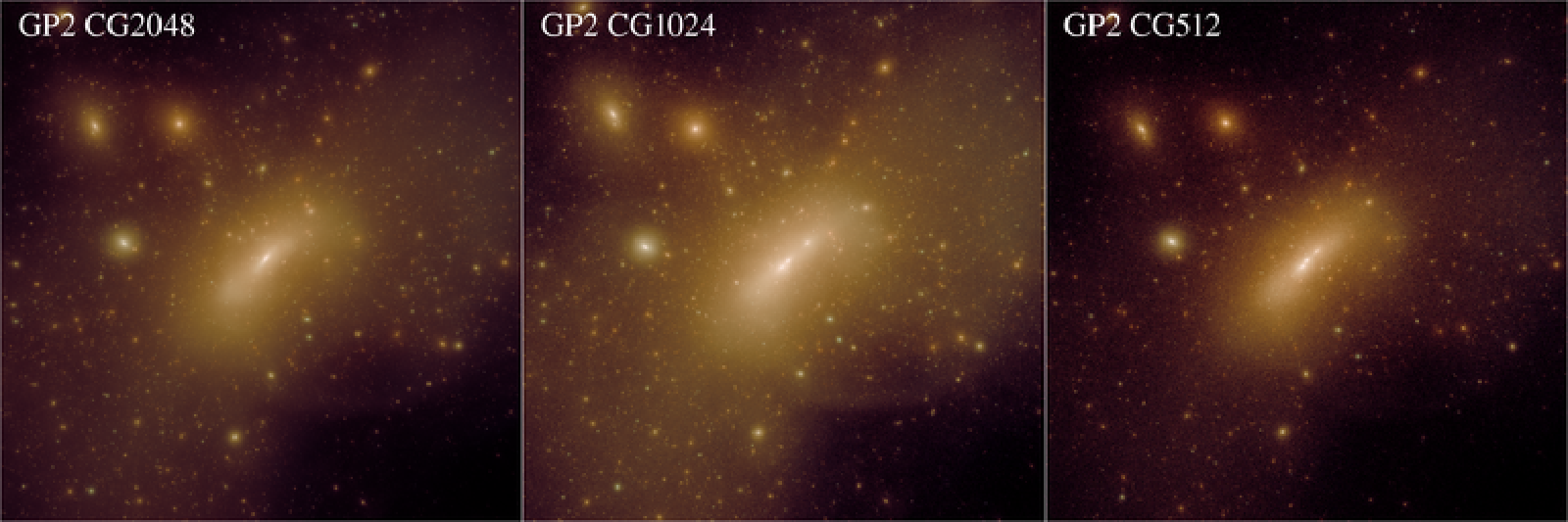}
\includegraphics[width=16cm]{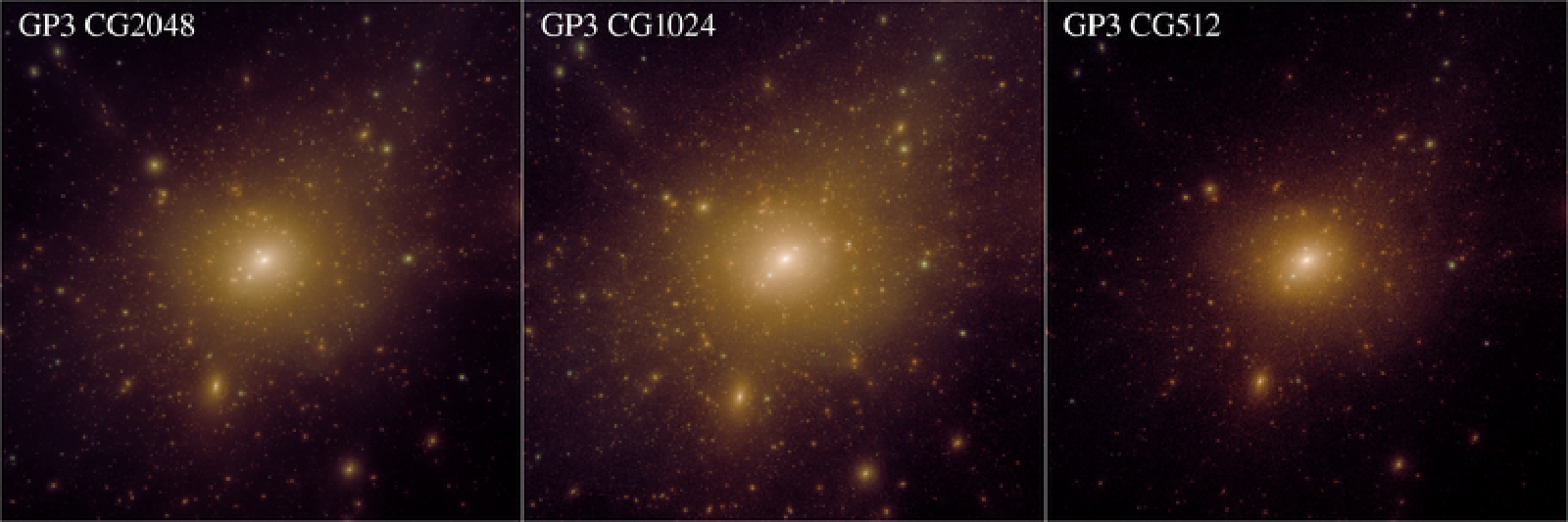}
\caption{
Projected density of dark matter at $z=0$. Each row shows one of 
the three most massive halos with mass decreasing from top to bottom. 
Columns show different resolution from highest (left) to lowest (right). 
}
\label{fig:res}
\end{figure*}

Figure \ref{fig:snapshot1} shows the snapshots at $z=0$. 
In Figure \ref{fig:snapshot2}, we also present
the time evolution of the whole box and that of the most massive halo.  
The three most massive halos in simulations with 
three different resolutions are shown in Figure \ref{fig:res}.
The positions of subhalos agree very well in three simulations.
Of course, there are some discrepancies near the centers of halos.  
In particular, whereas 
there is only one core in the center of the 
second massive halo (GP2) of CG2048, 
there are two cores in GP2 of CG1024 and CG512.  

The reason of this difference 
is that the formation history of this halo is rather violent. 
It experienced many mergers near $z=0$ in the center of the halo 
and is far from the relaxed state.
The difference of the accuracy of integration changed
the timescale of the mergers of the halos with 
three different resolutions.
At $z=0$, the halo GP2 has just completed the merger in the CG2048 run, 
whereas the same merger event is still on-going in CG1024 and CG512 runs. 
If we consider the
spherically averaged density profile of the halo, 
the difference becomes important (see Section \ref{sec:profile}).

\section{Results}\label{sec:result}

\subsection{Mass Function}\label{sec:massfunc}

\citet{Press1974} established a recipe to derive the
number of dark matter halos based on the hierarchical clustering model.
Since then, a number of analytic formulae for the mass function
have been proposed.
Many of them are designed to give a good agreement 
with results of high-resolution $N$-body simulations
\citep[e.g.,][and references therein]
{Sheth1999, Jenkins2001, Reed2003, Yahagi2004, Warren2006,
Tinker2008}.

These formulae can reproduce the mass function
between $10^{10} M_{\odot}$ and cluster scale very well.  
Here, we examine the mass function of mass 
below $10^{10} M_{\odot}$ down to $10^{7} M_{\odot}$.
The mass function of this range has been 
studied only in high redshift \citep[e.g.,][]{Reed2007, Lukic2007}.

\begin{figure*}
\centering
\includegraphics[width=5.9cm]{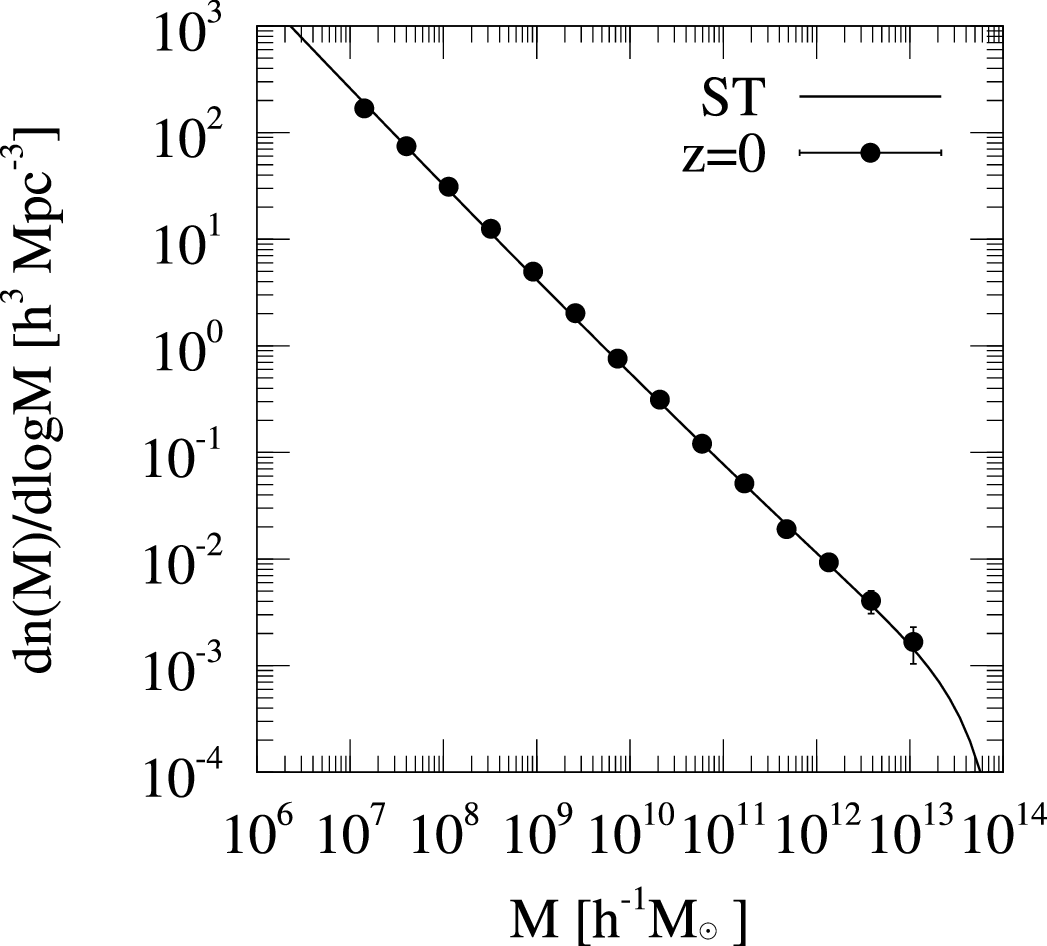}
\includegraphics[width=5.9cm]{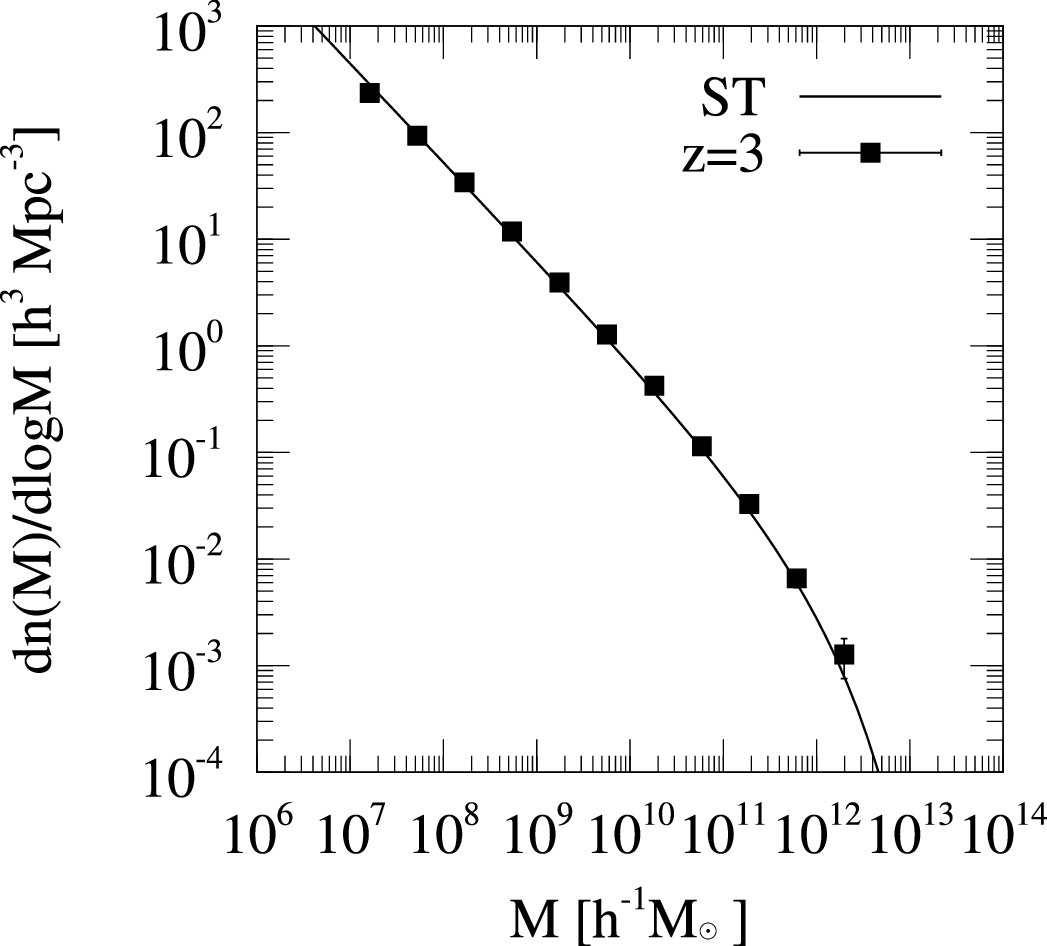}
\includegraphics[width=5.9cm]{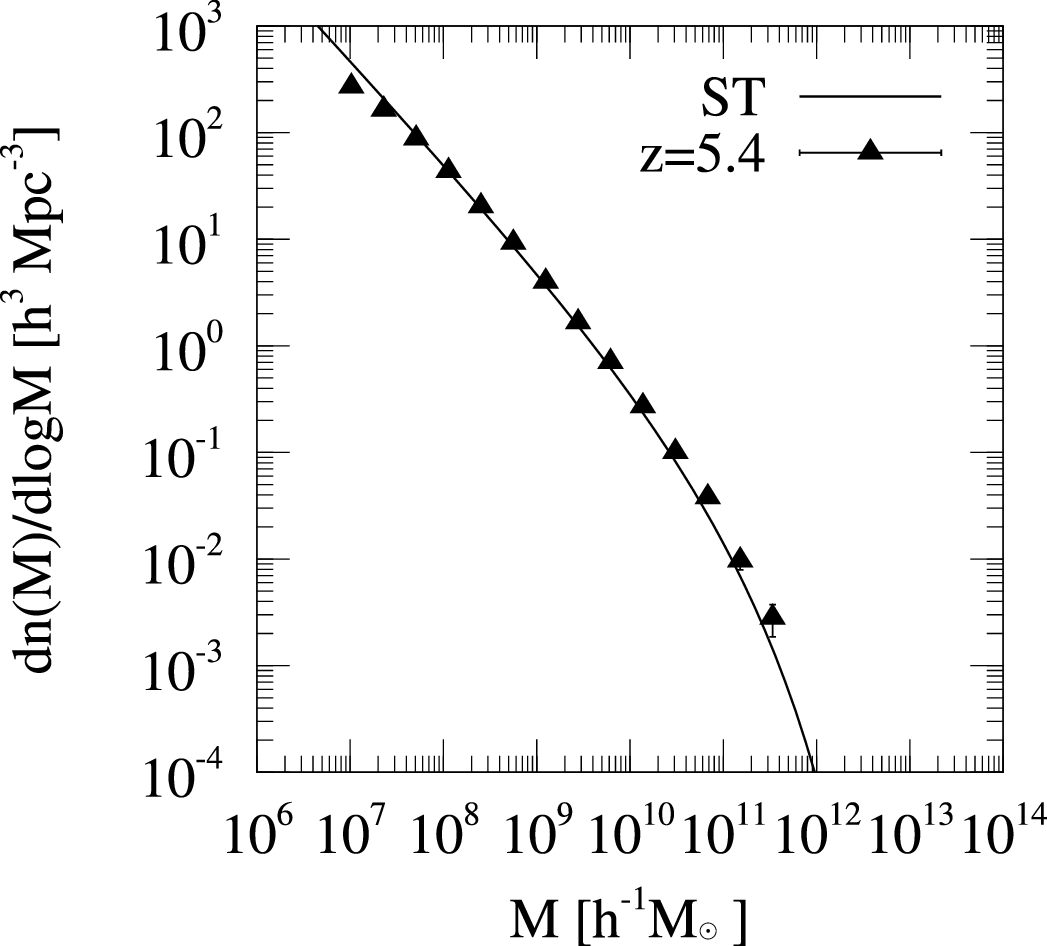}
\caption{
Mass function of our largest simulation (CG2048).
The results of $z=0.0$ (top-left), $z=3.0$ (top-right), and $z=5.4$ (bottom) are shown.
Solid curves are the \citet{Sheth1999} function.
Error bars are Poisson errors.
}
\label{fig:massfunc}
\end{figure*}

\begin{figure*}
\centering
\includegraphics[width=6cm]{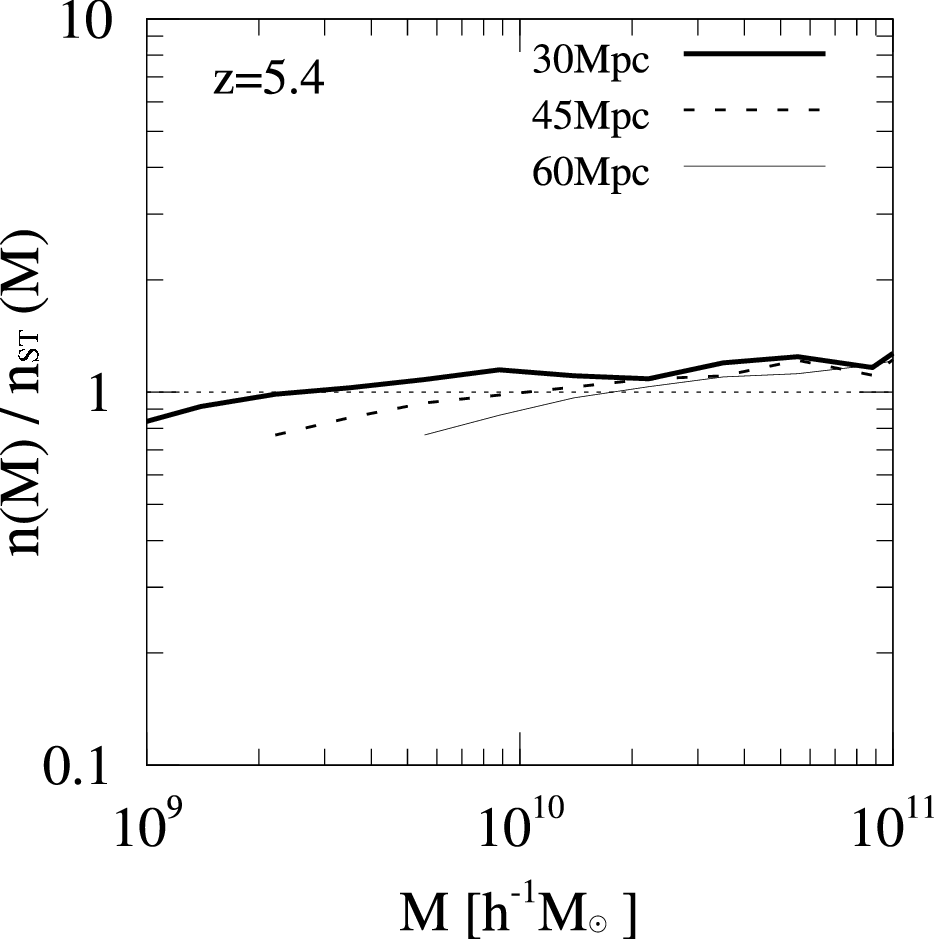}
\includegraphics[width=6cm]{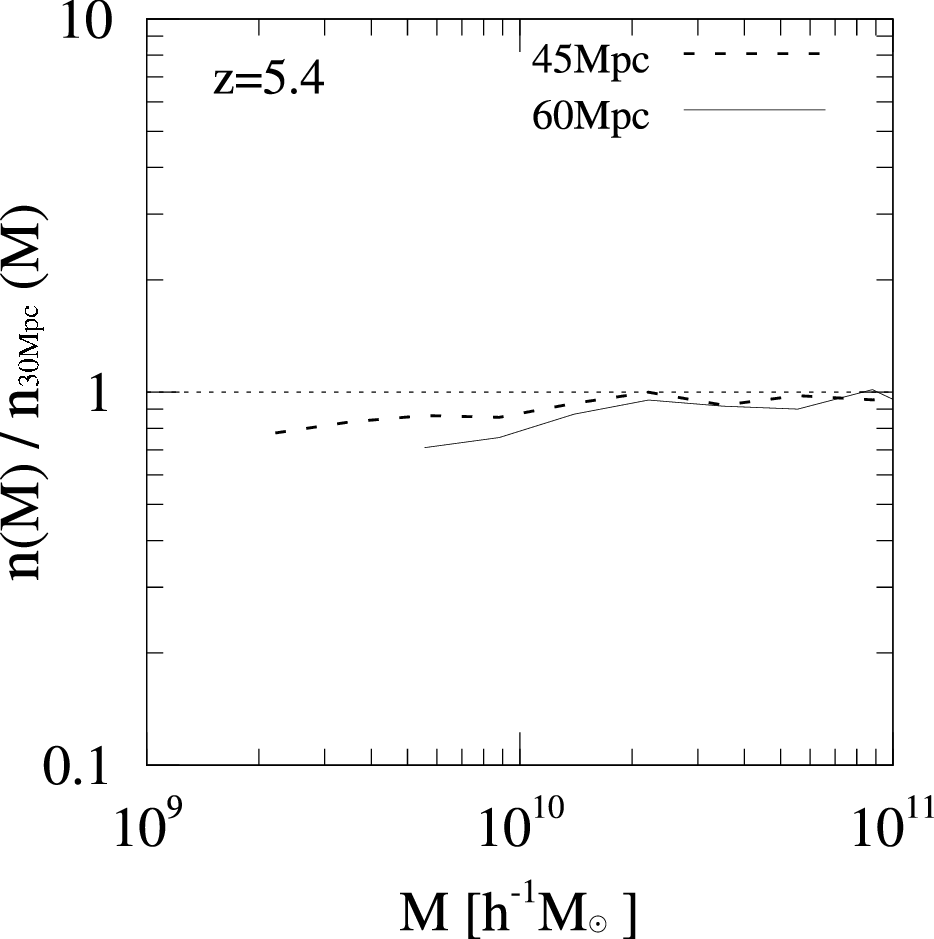}
\caption{
Left panel shows three mass functions at $z=5.4$
derived from $512^3$ simulations of 30, 45, and 60 Mpc boxes, 
relative to the \citet{Sheth1999} function.
Right panel shows mass functions of 45 and 60 Mpc boxes simulations,
 relative to that of the 30 Mpc box simulation.
}
\label{fig:massfunc2}
\end{figure*}

Figure \ref{fig:massfunc} shows the halo mass functions at three different
redshifts for CG2048 run and the prediction of 
Sheth \& Tormen formula \citep[ST,][]{Sheth1999}. 
The agreement is very good for the mass from
$\sim 10^7 M_{\odot}$ to $M=1.0 \times 10^{13} M_{\odot}$ at $z=0$. 
The difference is less than 10\% for 
$M=5.0 \times 10^7 M_{\odot}$ to $M=2.0 \times 10^{12} M_{\odot}$ at $z=0$,
$M=5.0 \times 10^7 M_{\odot}$ to $M=5.0 \times 10^{10} M_{\odot}$ at $z=3$, and
$M=8.0 \times 10^7 M_{\odot}$ to $M=4.0 \times 10^{9} M_{\odot}$ at $z=5.4$.

Our results imply that the mass function is well represented by the ST
function down to $10^7 M_{\odot}$.  However,
our simulations have a slightly larger number of halos than the number predicted
by the ST formula
in particular at the high-mass end of the $z=5.4$ mass function.
Note that
the finite volume of our simulation (the box length is 30Mpc) might
affect the mass function in some degrees.  
The absence of long-wavelength perturbations might increase 
the number of intermediate mass halos by about 10\% \citep{Bagla2006, Power2006}.
In order to test the effect of the box size, 
we performed additional simulations of 30, 45, and 60Mpc boxes with $512^3$ particles. 
The left panel of Figure \ref{fig:massfunc2} shows mass functions of
these simulations at $z=5.4$ 
relative to the ST formula.
The difference becomes larger as the halo mass and the box size increase. 
The right panel of Figure \ref{fig:massfunc2} shows 
mass functions relative to the 30Mpc simulation.
We can see that the number of halos of the 30Mpc box
simulation is systematically larger than those of the 45Mpc and 60Mpc box
simulations. The mass functions of the 45Mpc and 60Mpc box simulations
are well converged for halos larger than
 $2.0 \times 10^{10} M_{\odot}$, which is 
the limit of resolution for the 60Mpc box simulation. 
We can conclude that the larger number of halos seen in CG2048 
at the high-mass end is 
caused by the absence of long-wavelength perturbations.

\subsection{Density Structures of Most Massive Halos}\label{sec:profile}

Many groups have studied the density profile of dark matter halos using
high-resolution cosmological $N$-body simulations \citep[e.g.,][]{Navarro1997,
Fukushige1997, Moore1999, Ghigna2000, Jing2000a, Jing2000b,
Fukushige2001, Klypin2001, Taylor2001, Jing2002, Power2003, Fukushige2003,
Fukushige2004, Diemand2004b, Hayashi2004, Navarro2004, Diemand2005b,
Reed2005b, Kazantzidis2006, Merritt2006, Diemand2008, Gao2008, Stadel2009,
Navarro2010}.  In most of recent works, the slopes of radial
density profiles were around $-1$ in the inner region and around $-3$
in the outer region.  The slope of density became shallower as the radius
becomes smaller. Thus, the central slope is not described by any single power.
Furthermore, the density profile was not universal.  In
other words, the slope showed a significant halo-to-halo scatter.

\begin{figure}
\centering
\includegraphics[width=8cm]{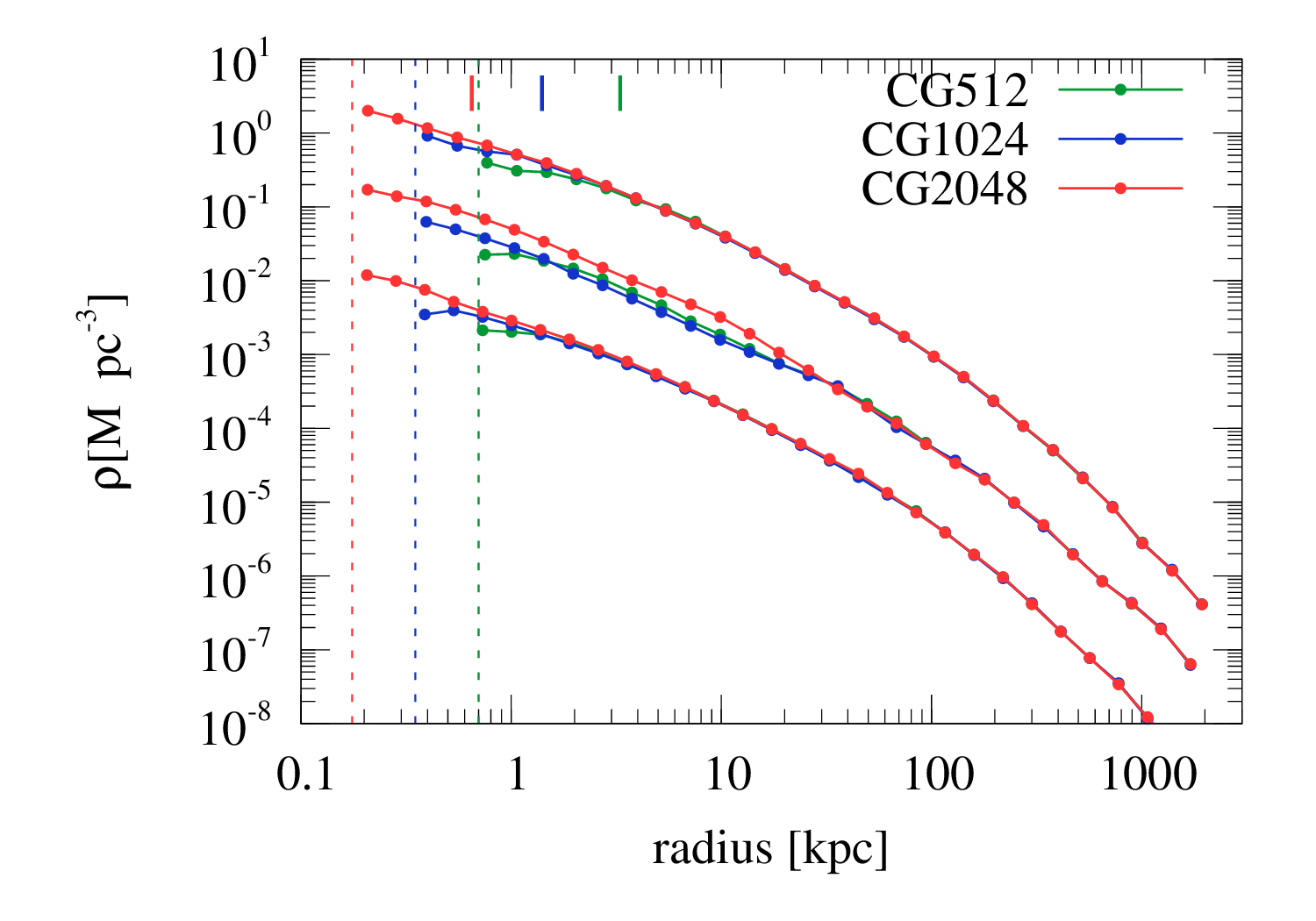}
\caption{
Spherically averaged radial density profiles of largest three halos at $z=0$. 
Two of three profiles (middle and bottom) are vertically shifted downward 
by 1 and 2 dex.
Vertical dashed lines show the softening length of three simulations.
Upside short vertical bars indicate the reliability limit 
of the most massive halo calculated using criterion
proposed by \citet{Fukushige2001} and \citet{Power2003}. 
The red, blue, and green correspond to the simulation 
CG2048, CG1024, and CG512.
}
\label{fig:density}
\end{figure}

\begin{figure}
\centering
\includegraphics[width=8cm]{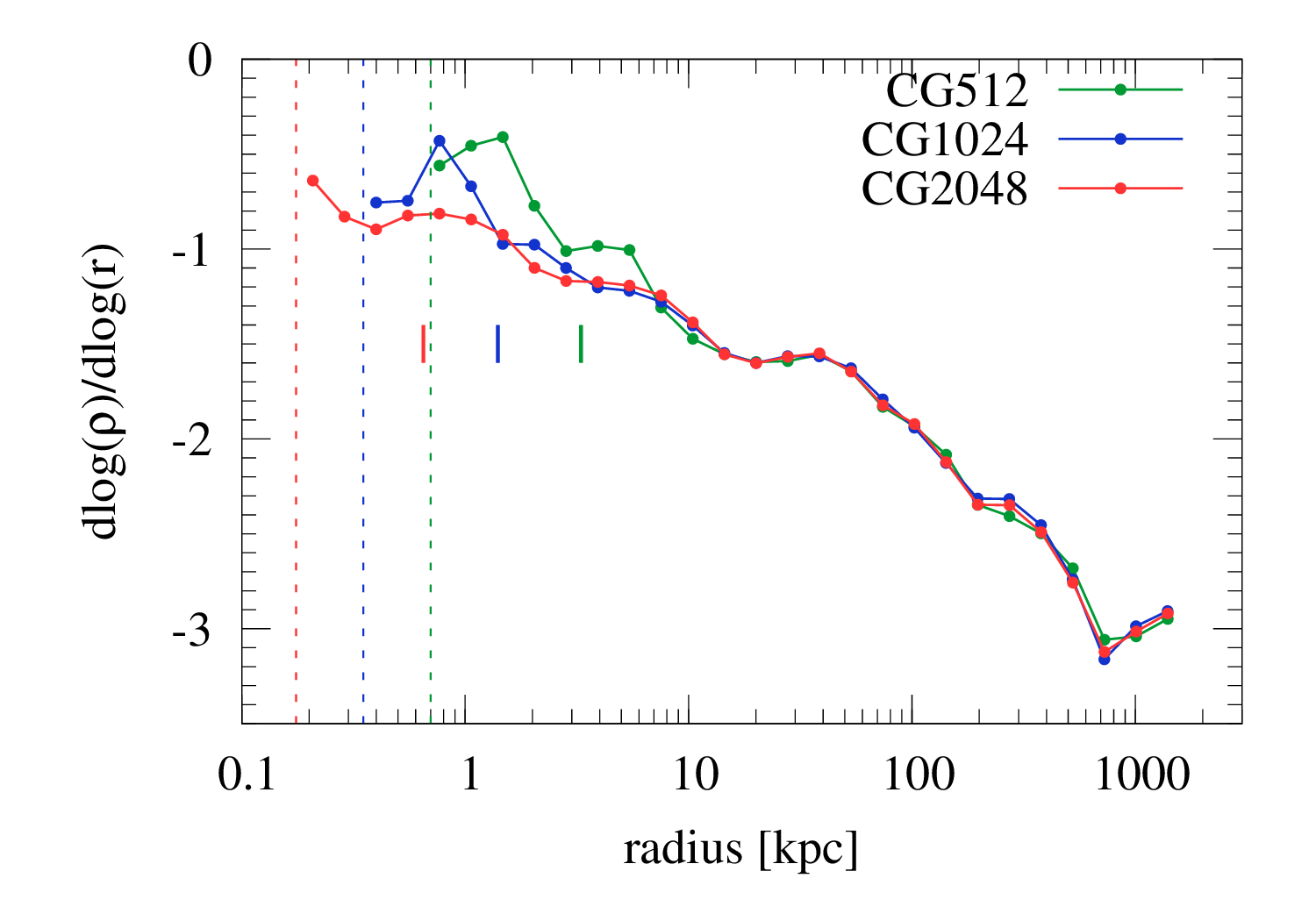}
\includegraphics[width=8cm]{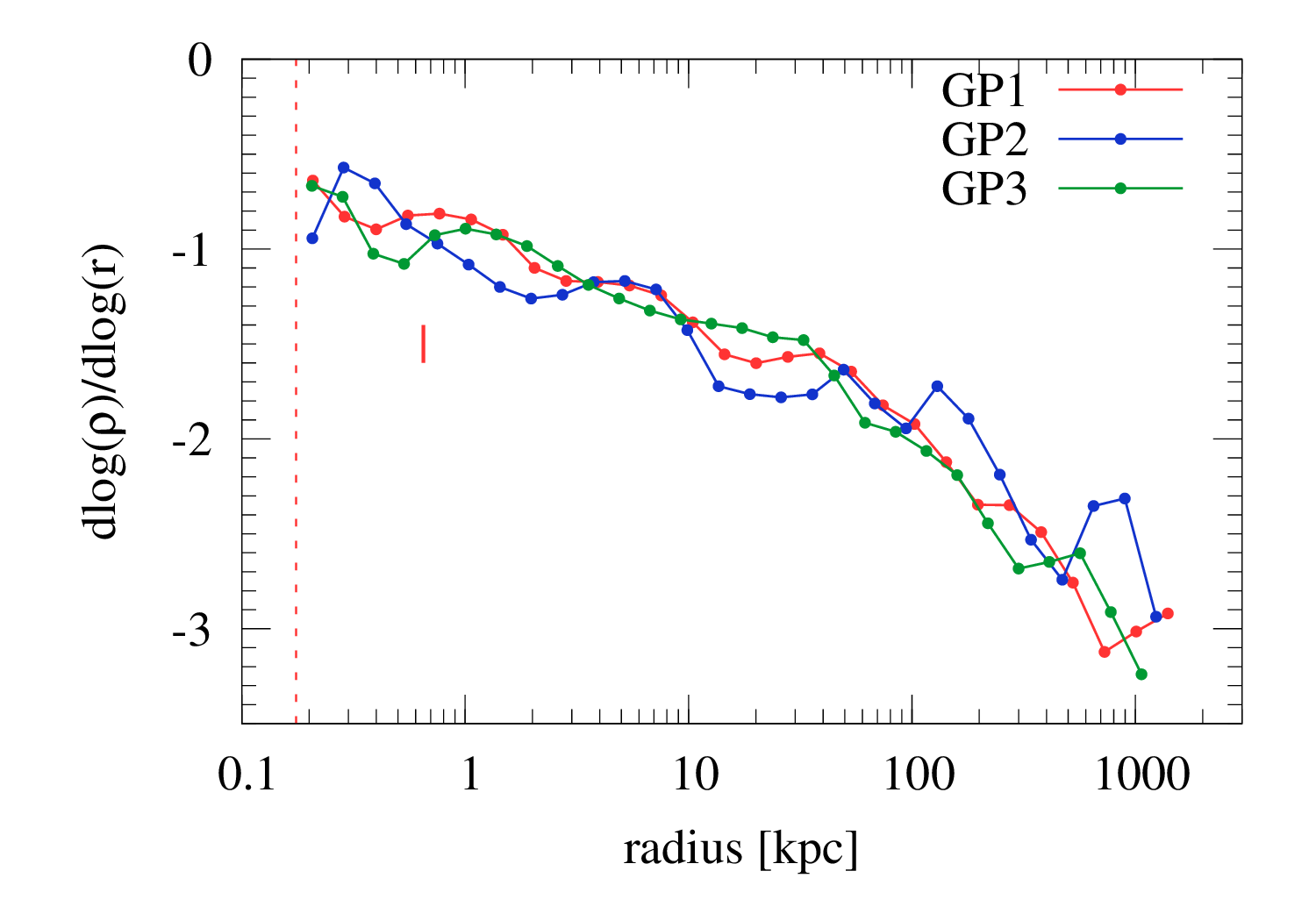}
\caption{
Slopes of radial density profiles of largest three halos at $z=0$. 
Left panel shows those of the largest halo for three different resolutions. 
Right panel shows those of the largest three halos for the largest simulation (CG2048).
}
\label{fig:slope}
\end{figure}

Recent studies \citep{Stadel2009, Navarro2010} based on high-resolution
simulations of galactic halos showed that the slopes of density
became less than $-1$ at the radius $0.001$ times the virial radius of the halo 
as predicted by early works \citep[e.g.][]{Graham2006}.
Einasto profile showed better agreement than the NFW profile \citep{Navarro1997} 
which has been widely used for modeling dark matter halos
because of its simplicity.

Almost all recent high-resolution simulations of single halos
used galaxy-sized halos. 
Therefore, little is known if these finding can be applied 
to halos of different masses.
Here, we present the density
profiles of three most massive halos in our simulation.
These halos are galactic group-sized ones, 
with the mass of 5.24, 3.58, and 2.25 $\times 10^{13} M_{\odot}$. 
They contain 408, 279, and 176 million particles.

The top panel of Figure \ref{fig:density} shows the spherically averaged
density profiles of these halos at $z=0$.  We can see that the
results of three simulations with different resolution are indistinguishable
for radii larger than the reliability limits, except for the second massive halo. 
We calculated the reliability limits using criterion 
proposed by \citet{Fukushige2001} and \citet{Power2003}.
We cannot ignore the effects of the local two-body relaxation 
for radii smaller than these limits.
As can be seen in Figure \ref{fig:res}, the slight difference of the
merging epoch of the central cores caused this difference.

The slopes of density profiles
become gradually shallower as the radius becomes smaller.  The left panel of
Figure \ref{fig:slope} shows the slopes of density profiles of the most
massive halo.  As in the case of the density profile itself, 
the slopes also agree well with each other. The right panel of
Figure \ref{fig:slope} shows the slopes of the three most massive halos
in CG2048 run. 

These profiles are significantly different from those
  of galactic halos in recent other high-resolution simulations, even
  if the halo mass is scaled to be the same.  The mass of the halos of
  Aquarius simulation \citep{Springel2008} or GHALO simulation
  \citep{Stadel2009} is $\sim 10^{12}M_{\odot}$, which is an order of
  magnitude smaller than our three halos.  
The slope at $0.001R_{\rm vir}$ is $-0.9 \sim -1.0$ for our three halos.  
This value is in
excellent agreement with the result of both simulations.  Both of them
gave the slope $-1.0$ for $r=0.001r_{200}$.  this agreement does not
mean the density profile obtained by these simulation and those by our
simulation are identical.  The concentration parameter, which we
define here as $c_{\rm vmax} = R_{\rm vir} / R_{\rm vmax}$, is 4.8,
where $R_{\rm vir}$ and $R_{\rm vmax}$ are the halo virial radius and
the radius of the maximum rotational velocity.  This value is
significantly smaller than that of Aquarius A-1 halo.  Thus, the
Aquarius halo is significantly more centrally concentrated, and yet
the slope at $r=0.001R_{\rm vir}$ is the same.  Thus the rate of the
shallowing of the slope is somewhat faster for the Aquarius halo than
for our CG2048 halos.  Most likely, this difference is due to the
difference in the mass of the halo.

\subsection{Concentration Distributions}
The concentration parameter has been widely used to describe 
the internal structure of halos
since it is tightly correlated with the formation epoch \citep{Wechsler2002}.
Usually, the concentration is parameterized assuming that 
the density profiles of halos can be fitted by the NFW profile \citep{Navarro1997}, 
\begin{eqnarray}
\rho(r) = \frac{\rm \rho_0}{(r/r_{\rm s})(1+r/r_{\rm s})^2},
\end{eqnarray}
where $\rho_0$ is a characteristic density and $r_s$ is a scale radius.
The concentration $c_{\rm NFW} = R/r_s$ is widely used 
\citep[e.g.,][]{Bullock2001, Zhao2003, Maccio2007, 
Neto2007, Maccio2008, Zhao2009, Munoz2010}.
It is known that $c_{\rm NFW}$ depends weakly on the halo mass. 
Halos with higher mass have smaller concentration, since the average density 
of a halo reflects the cosmic density at its formation time.
The dependence is weaker for higher redshift \citep{Zhao2003}.

\begin{figure}
\centering \includegraphics[width=8cm]{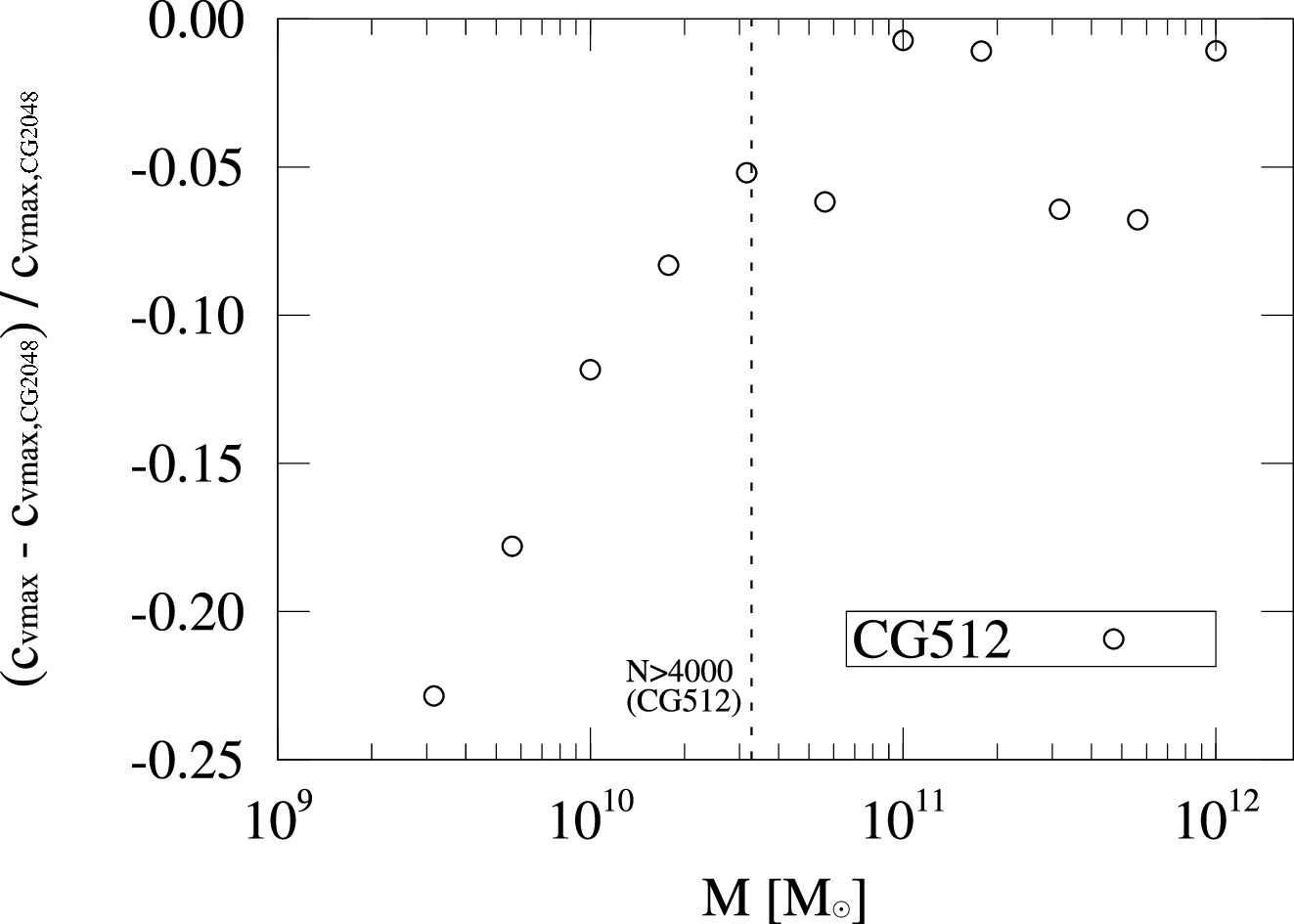} 
\caption{
Residuals of concentration $c_{\rm vmax} = R_{\rm vir}/R_{\rm vmax}$ 
from the largest simulation (CG2048) to the lower resolution simulation (CG512).
} 
\label{fig:m-cvmax_res}
\end{figure}

\begin{figure}
\centering \includegraphics[width=8cm]{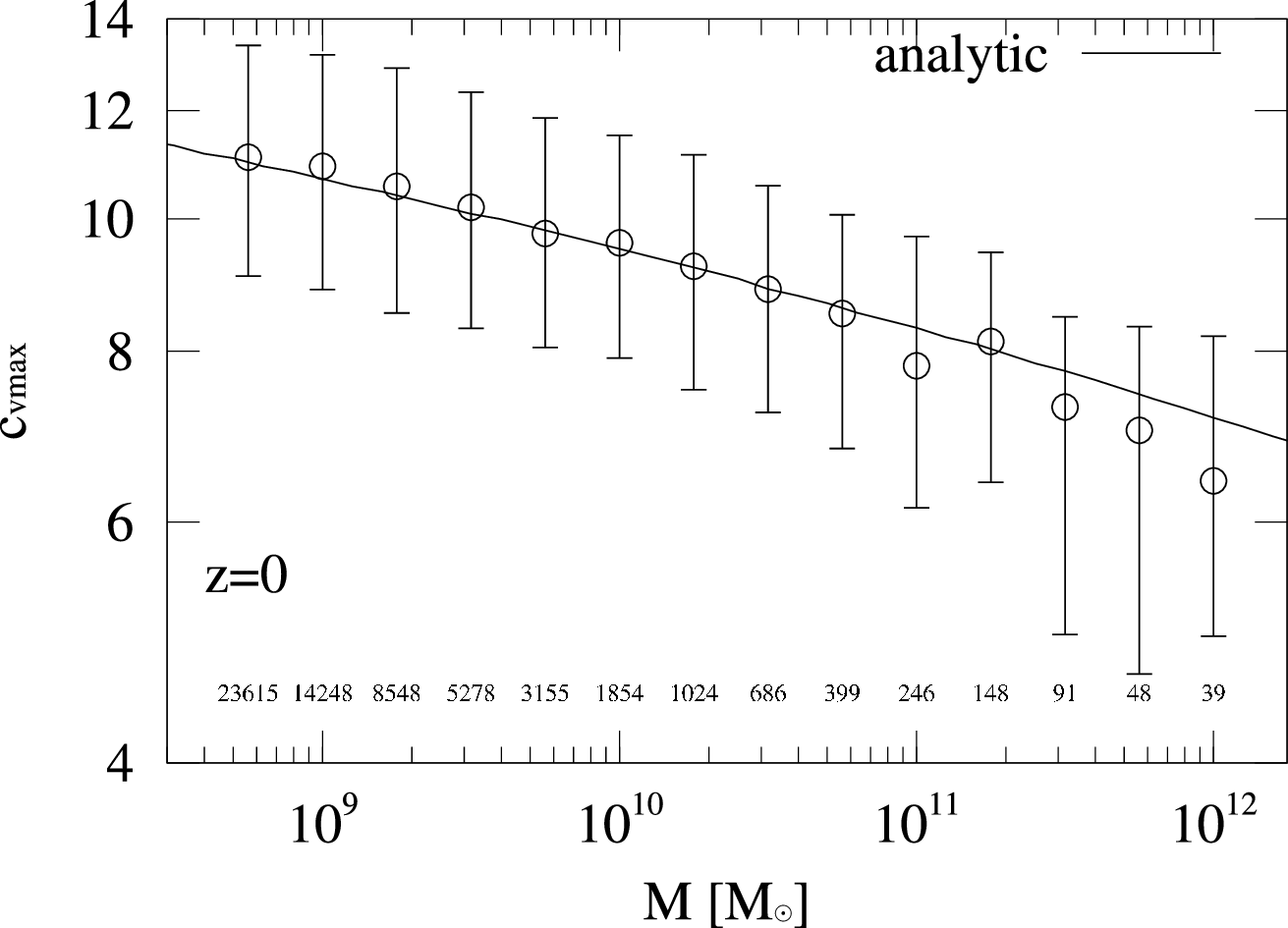} 
\caption{
Concentration plotted against the halo
virial mass $M$ at $z=0$.  Circles show the median value on each bin.
Whiskers are the first and third quantiles.  The number of halos on each
bin is shown below circles. 
Thick solid line shows the result from an analytical model. 
} \label{fig:m-cvmax}
\end{figure}

The concentration based on the NFW profile is
affected by fitting ranges and resolution \citep{Neto2007}. 
Furthermore, recent high-resolution simulations showed 
that the density profile is significantly different from the NFW profile 
\citep[][also see Section \ref{sec:profile}]{Stadel2009, Navarro2010}. 
Thus, the use of $c_{\rm nfw}$ might cause some systematic bias
\citep{Gao2008, Reed2011}.

We use the concentration $c_{\rm vmax}$ defined in Section \ref{sec:profile}, 
which is a simpler quantity to measure the concentration. 
Note that $R_{\rm vmax}$ can be 
easily determined directly from spherically averaged mass distribution
without the need of any fitting formulae.
If the density profile is represented by
the NFW profile, either concentration can be converted to the other.

First, we determine the minimum number of particles 
in a halo necessary to reliably determine the concentration.
Figure \ref{fig:m-cvmax_res} shows the normalized difference of 
average concentration between the G2048 run and the CG512 run 
as the function of halo mass. 
We can see that the difference is $\sim 0.05$ 
for the halo mass larger than $3.0 \times 10^{10} M_{\odot}$. 
For halo mass less than $3.0 \times 10^{10} M_{\odot}$, 
the difference is larger. 
In the CG512 run, a halo of mass $3.0 \times 10^{10} M_{\odot}$
contains $\sim 4000$ particles.
So we conclude that we need $\sim 4000$ particles 
to reliably determine the concentration.
For the CG2048 run, the reliability limit is $5.0 \times 10^8 M_{\odot}$.

\begin{figure*}
\centering
\includegraphics[width=5.9cm]{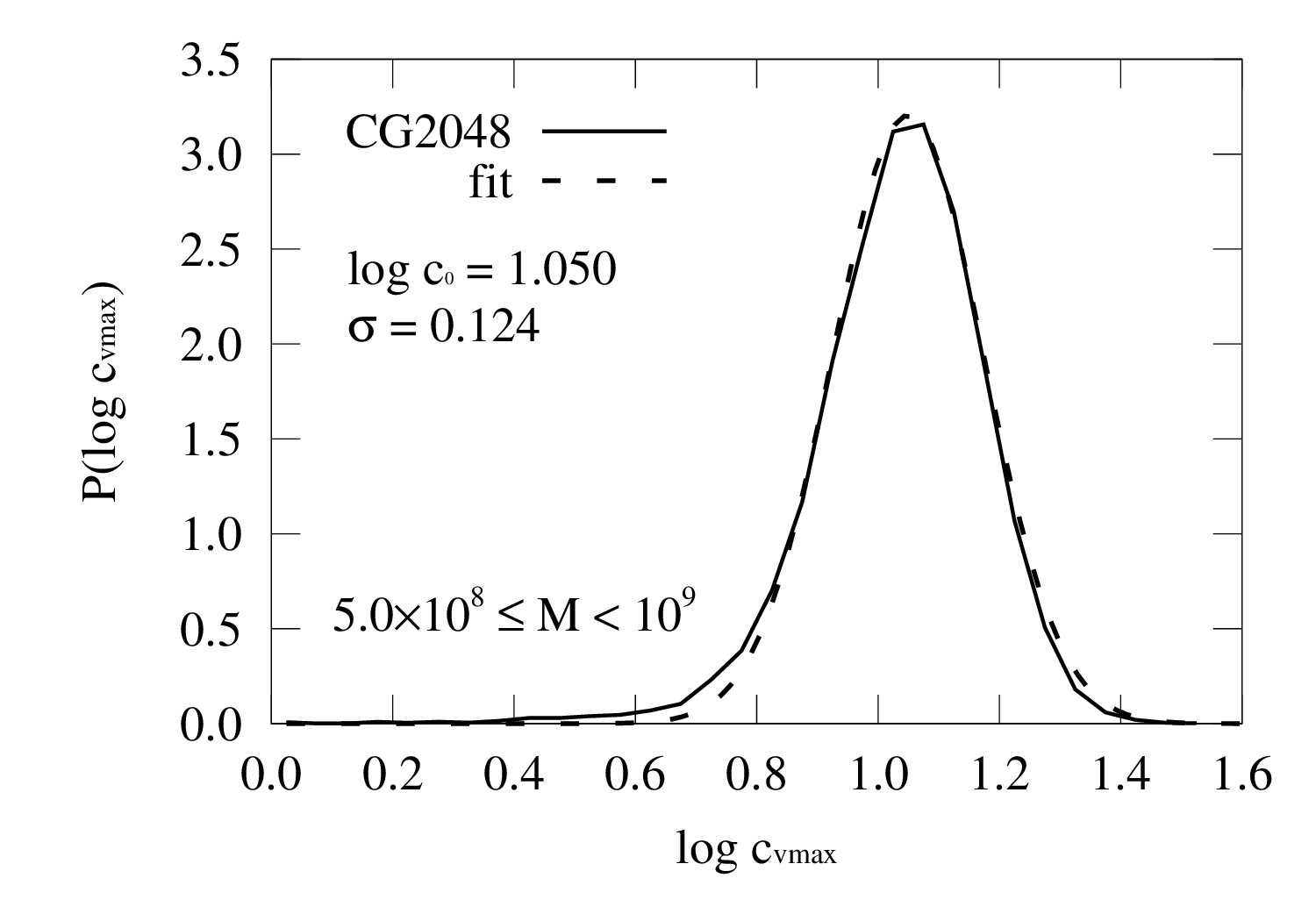}
\includegraphics[width=5.9cm]{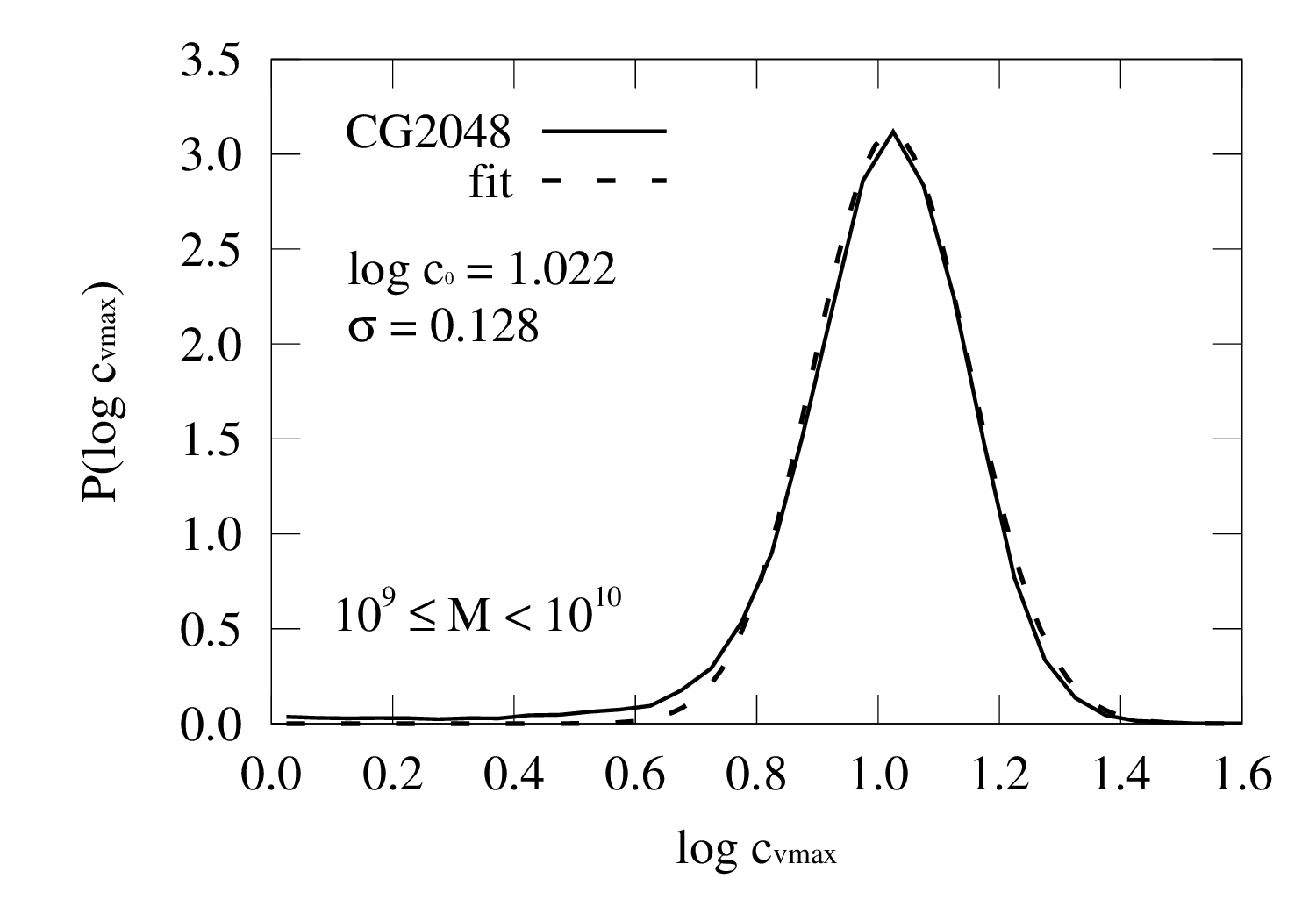}
\includegraphics[width=5.9cm]{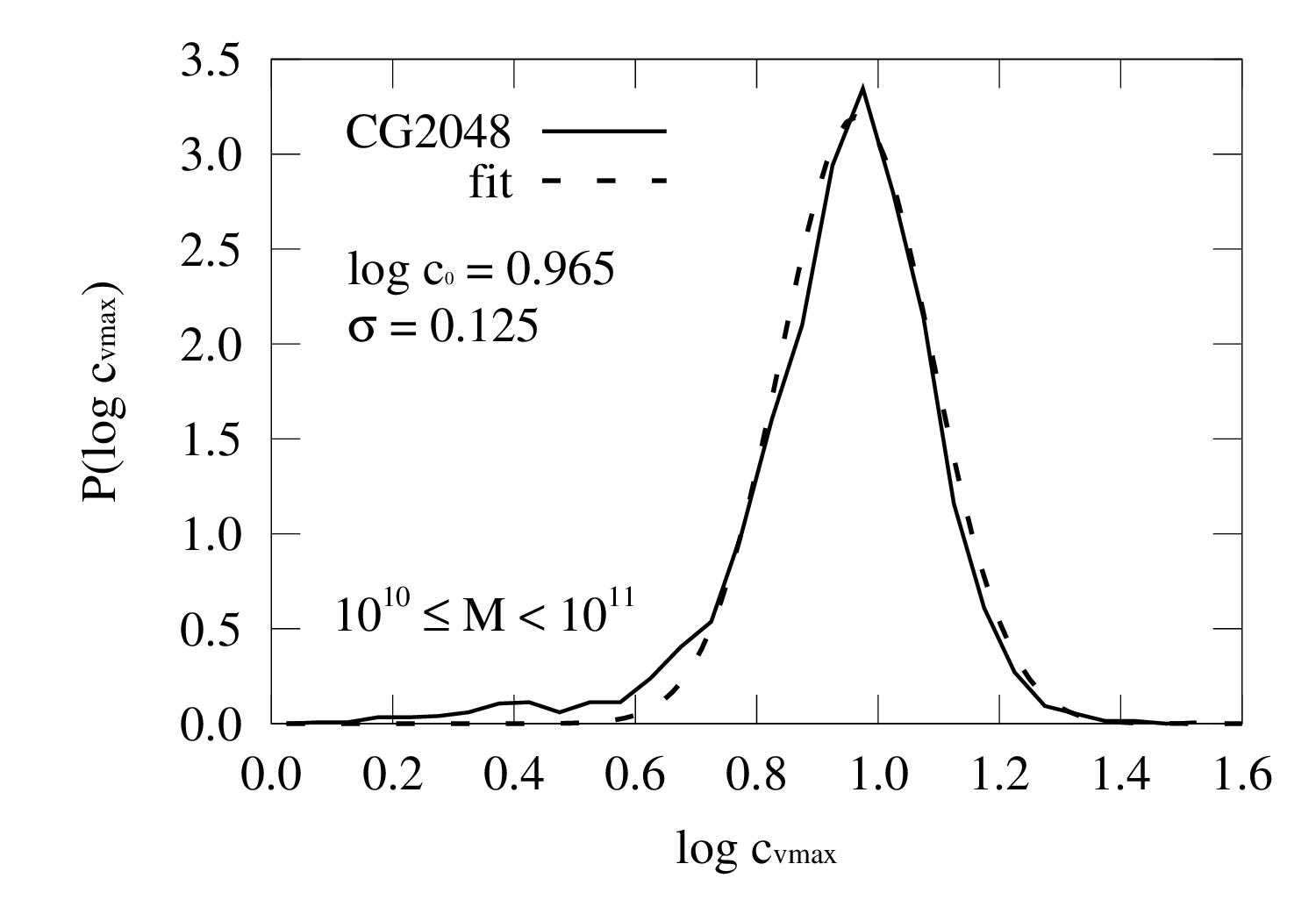}
\caption{
Probability distribution functions of the concentration at $z=0$. 
These panels show the results of different mass ranges.
Dashed curves are the best fits of the log-normal distribution.
}
\label{fig:cvmax_pba}
\end{figure*}

Figure \ref{fig:m-cvmax} shows the median, and first and third quantiles of
the concentration as a function of the virial mass of the halo.  
We can see a clear correlation between
the halo mass and the concentration.
Apparently, the dependence is weaker for smaller mass.
Therefore, the fitting functions with a single power
\citep[e.g.,][]{Bullock2001b, Neto2007, Maccio2007, Klypin2011} cannot be
used for halos of the size of dwarf galaxies.

Theoretically, the concentration of a halo reflects the cosmic density
at the formation time of the halo \citep{Bullock2001b}.  
The concentrations of halos formed earlier 
are higher than that of halos formed later.
However, the
dependence should be weak for small halos
since the dependence of the formation epoch to the halo mass 
is small for small (smaller than $10^8 M_{\odot}$) halos.
The slope of the power spectrum of initial density 
fluctuations approaches to $-3$ for small mass limit.

In Figure \ref{fig:m-cvmax}, we also plot an analytical prediction 
of the mass$-$concentration relation,
obtained by the method used in \citet{Navarro1997} 
assuming that all halos have the NFW profile.
The formation redshift $z_{\rm f}$ of halos with the mass $M$ is defined 
as the epoch at which progenitors with the mass larger than $fM$  
first contained the half of the mass $M$. 
It is estimated by using 
the Press Schechter formalism \citep[e.g.,][]{Lacey1993}, 
\begin{eqnarray}
{\rm erfc} \left\{ \frac{\delta_{\rm crit}(z_{\rm f}) - \delta_{\rm crit}(0)}
{\sqrt{2\left[ \sigma_0^2(fM) - \sigma_0^2(M) \right]}} \right\} = \frac{1}{2}, \label{eq:zf}
\end{eqnarray}
where $\delta_{\rm crit}(z)$ is the critical overdensity 
for the spherical collapse at $z_{\rm f}$, and 
$\sigma_0^2(M)$ is the variance of the density fluctuation at $z=0$ 
smoothed by a top-hat filter on a mass scale of $M$. 
Here, we used $f=0.01$.
The characteristic density $\rm \rho_0$ of a halo should reflect the cosmic density 
at the formation time. Thus, we assume
\begin{eqnarray}
\rho_0 = \rho_{\rm norm} \left( 1 + z_{\rm f} \right)^3, \label{eq:rho}
\end{eqnarray}
where $\rho_{\rm norm}$ is chosen to fit the simulation results. 
The mass of a halo with the NFW profile is given by
\begin{eqnarray}
M = 4 \pi \rho_0 r_{\rm s}^3 \left[ \ln(1+c) - c/(1+c) \right]. \label{eq:mass}
\end{eqnarray}
The mass and concentration at $z=0$ are related to each other by 
\begin{eqnarray}
M = \frac{4}{3} \pi R_{\rm vir}^3 \Delta(0) \rho_{\rm crit}
= \frac{4}{3} \pi r_{\rm s}^3 c^3 \Delta(0) \rho_{\rm crit}, \label{eq:rad}
\end{eqnarray}
where $\rho_{\rm crit}$ is the critical density. 
From eEuations (\ref{eq:overdensity}), (\ref{eq:zf}), (\ref{eq:rho}), 
(\ref{eq:mass}), and (\ref{eq:rad}), 
we can analytically estimate the concentration of halos with the mass $M$.

As mentioned by \citet{Lacey1993}, 
the estimated formation epoch obtained using Equation (\ref{eq:zf})
is not necessarily correct.
This is because the formation time defined here corresponds to the epoch
at which one of progenitors has a mass larger than $fM$.
This does not mean that the main progenitor has this mass.
Nevertheless, as seen in Figure \ref{fig:m-cvmax}, 
the analytical prediction based on Equation (\ref{eq:zf}) shows a very good agreement 
with the result from CG2048 run for halos with mass smaller than $10^{11} M_{\odot}$.
For halos with the mass larger than $10^{11} M_{\odot}$, 
the difference between CG2048 results and analytical ones 
are relatively large. However, these halos are rare objects in CG2048 run, 
and the fact might affect the results in some degrees.
We can conclude that the shallowing slope of the mass$-$concentration relation 
naturally emerges from the nature of the power spectrum 
of initial density fluctuations.

The slope is slightly shallower than that of $c_{\rm NFW}$ for larger
halos.  For the case of $c_{\rm NFW}$, the slope is around $-0.10$ for
relaxed halos and $-0.11$ for all halos \citep{Neto2007, Maccio2007}. 
On the other hand, for the CG2048 simulation, the slope is around $-0.07$ 
for halos with the mass $10^{10}M_{\odot}$, 
and $-0.06$ for halos with the mass $10^{9}M_{\odot}$.
Note that one overestimates the central density of halos
if one estimates the concentration of dwarf-sized halos 
by extrapolating the mass$-$concentration relation 
of galaxy or cluster-sized halos.

Figure \ref{fig:cvmax_pba} shows the probability distribution functions 
of the concentration parameter at $z=0$ in three different mass ranges. 
Both shapes are well fitted by the log-normal distributions,
\begin{eqnarray}
P(\log c_{\rm vmax}) = \frac{1}{\sqrt{2\pi} \sigma} 
\exp \left(-\frac{\log^2{(c_{\rm vmax}/c_0)}}{2\sigma^2} \right).
\end{eqnarray}

We find $\log c_0 = 1.050, \sigma = 0.124$ 
for halos with the mass of $5.0 \times 10^8 M_{\odot} \le M < 10^9 M_{\odot}$, 
$\log c_0 = 1.022, \sigma = 0.128$
for halos with the mass of $10^9 M_{\odot} \le M < 10^{10} M_{\odot}$, and
$\log c_0 = 0.965, \sigma = 0.125$
for halos with the mass of $10^{10} M_{\odot} \le M < 10^{11} M_{\odot}$.

\subsection{Spin Distributions}

The dimensionless spin parameter is a good parameter 
to quantify the rotation of a halo. One often uses the spin parameter 
defined in \citet{Bullock2001b}, 
\begin{eqnarray}
\lambda = \frac{J}{\sqrt{2}MVR},
\end{eqnarray}
where $M$, $R$, $V$, and $J$ are the virial mass of the halo, radius, 
rotational velocity at $R$, and total angular momentum inside $R$.

The distribution, the dependence on the halo mass, and the evolution have been
studied by a number of works \citep[e.g.,][]{Bullock2001b, Bailin2005, Bett2007,
Maccio2007, Knebe2008, Maccio2008, Delogu2010, Munoz2010, Wang2011}.  
The spin of galaxy-sized halos is well studied by using 
the results of sufficient resolution simulations.
However, we do not understand those of dwarf-galaxy-sized halos.
The spin distribution of those halos at only high redshifts is studied by 
the result of  high-resolution simulation \citep{Knebe2008}.  
Here, we extend the spin distributions at $z=0$ to dwarf-galaxy-sized 
halos (down to $10^8 M_{\odot}$) in the same way as the concentration.

\begin{figure}
\centering \includegraphics[width=8cm]{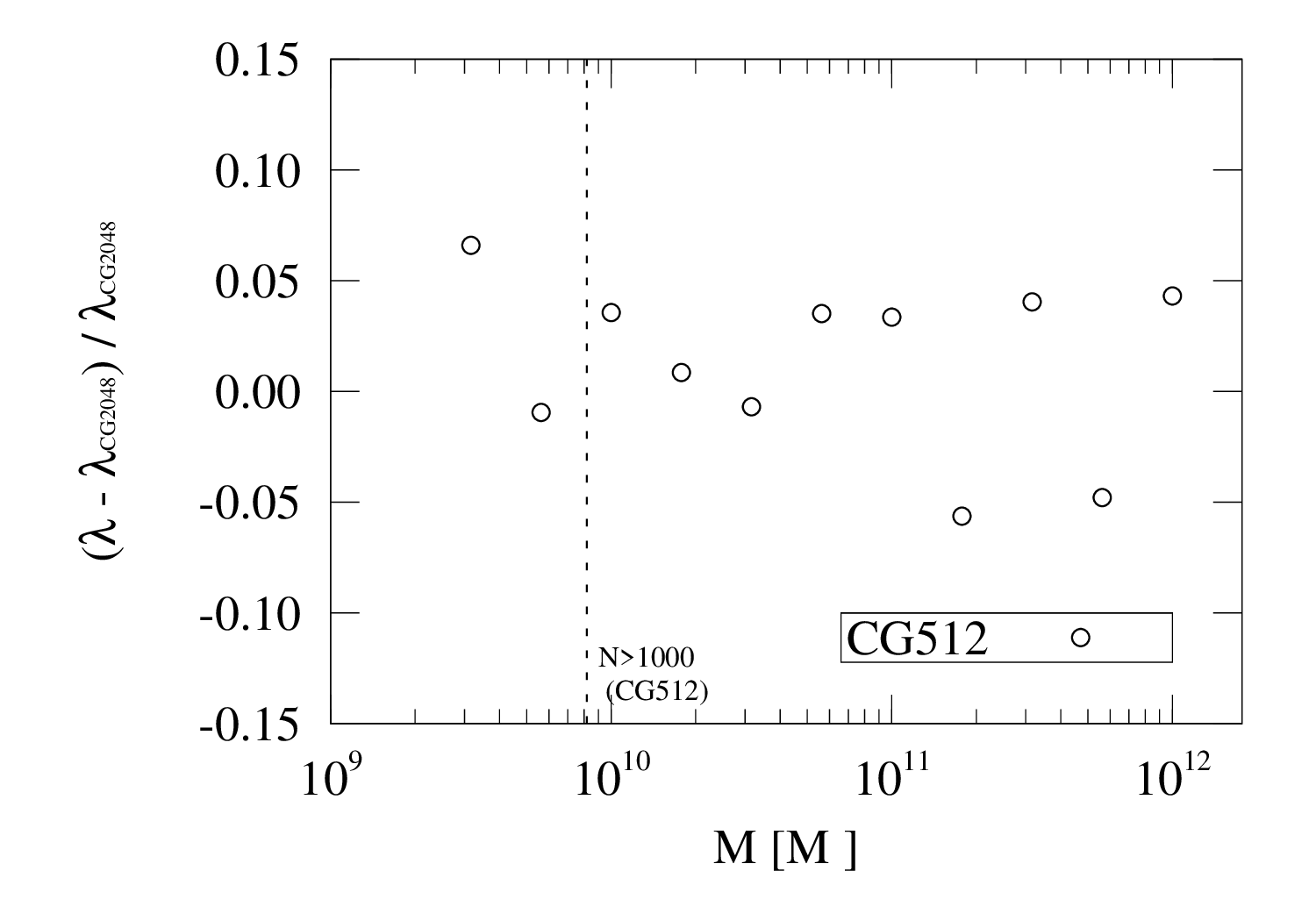} 
\caption{
Residuals of spin from the largest simulation (CG2048) to lower
resolution simulation (CG512).  } 
\label{fig:m-spin_res}
\end{figure}

\begin{figure}
\centering
\includegraphics[width=8cm]{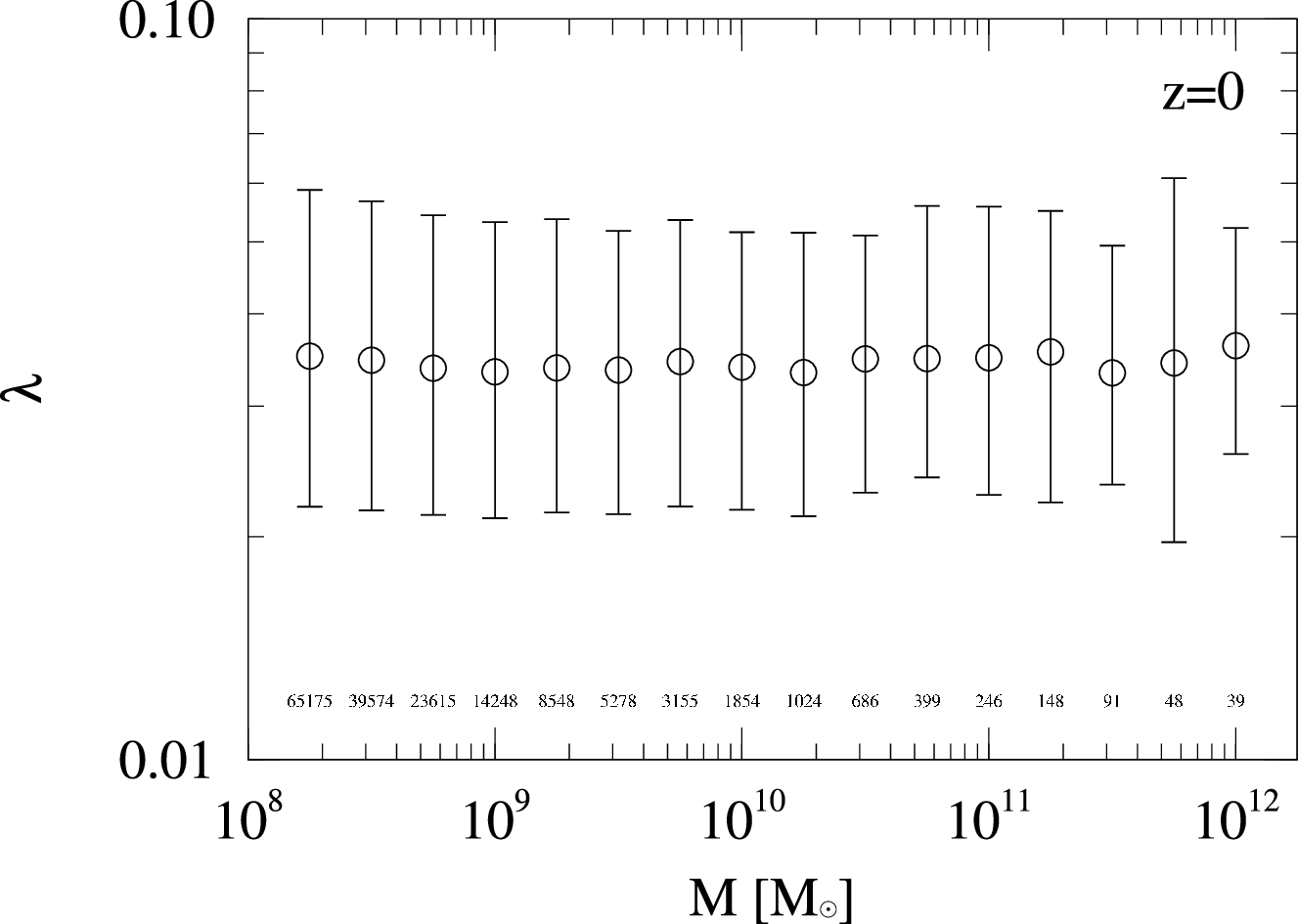}
\caption{
Spin parameter $\lambda$ plotted against the halo virial mass $M$ at $z=0$. 
Circles show the median value on each bin. 
Whiskers are the first and third quantiles. 
The number of halos on each bin is shown below circles.
}
\label{fig:m-spin}
\end{figure}

\begin{figure*}
\centering
\includegraphics[width=5.9cm]{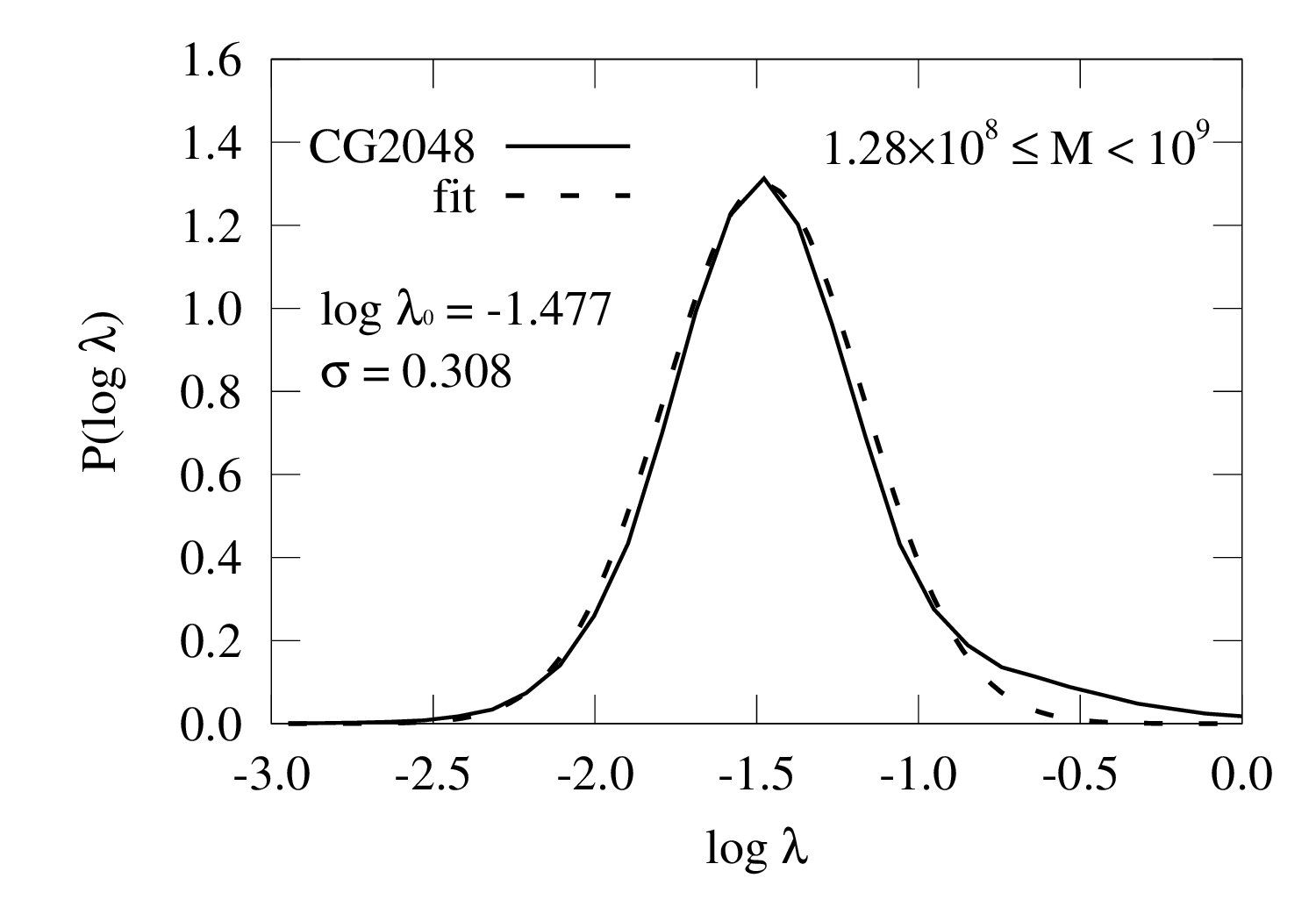}
\includegraphics[width=5.9cm]{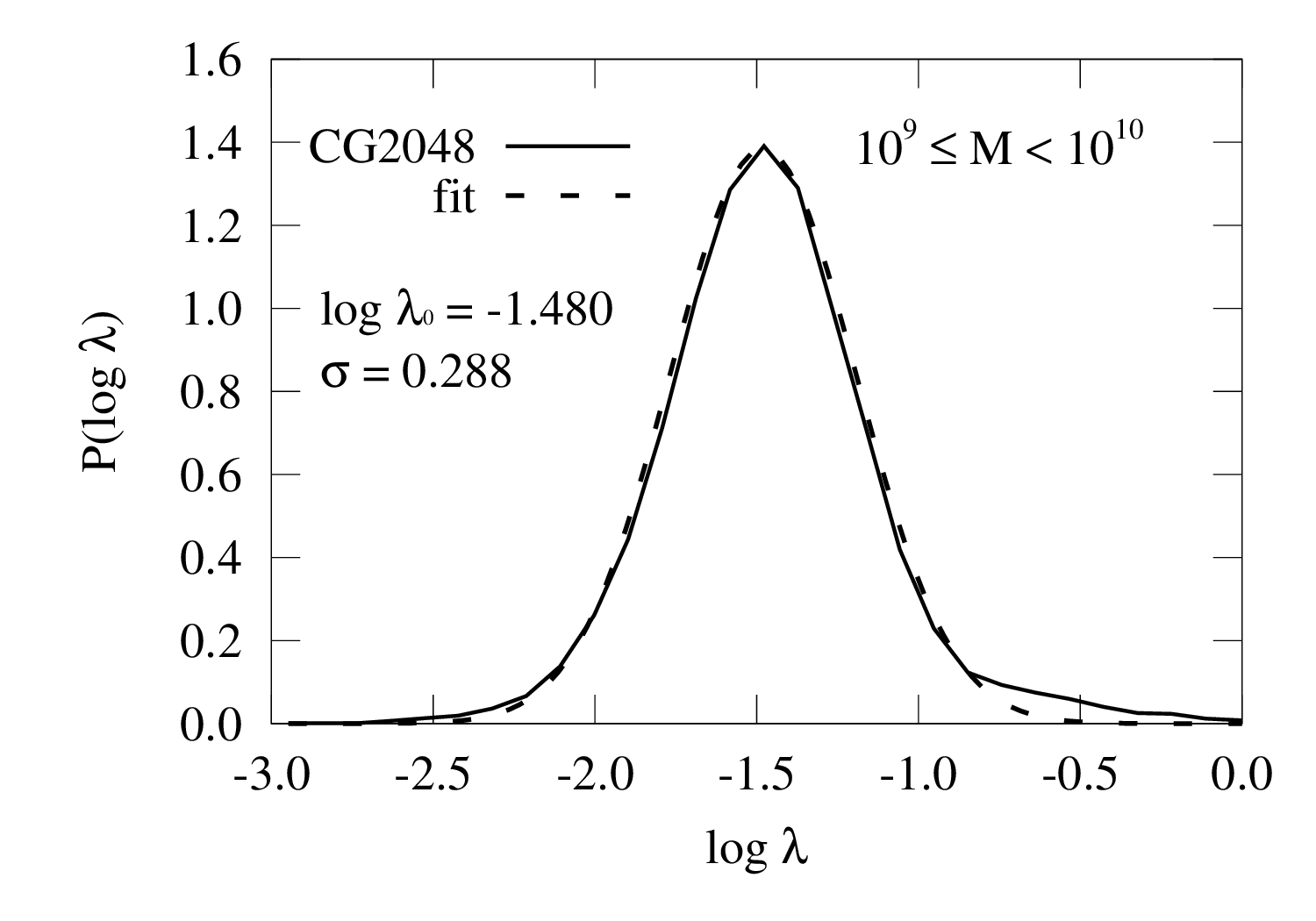}
\includegraphics[width=5.9cm]{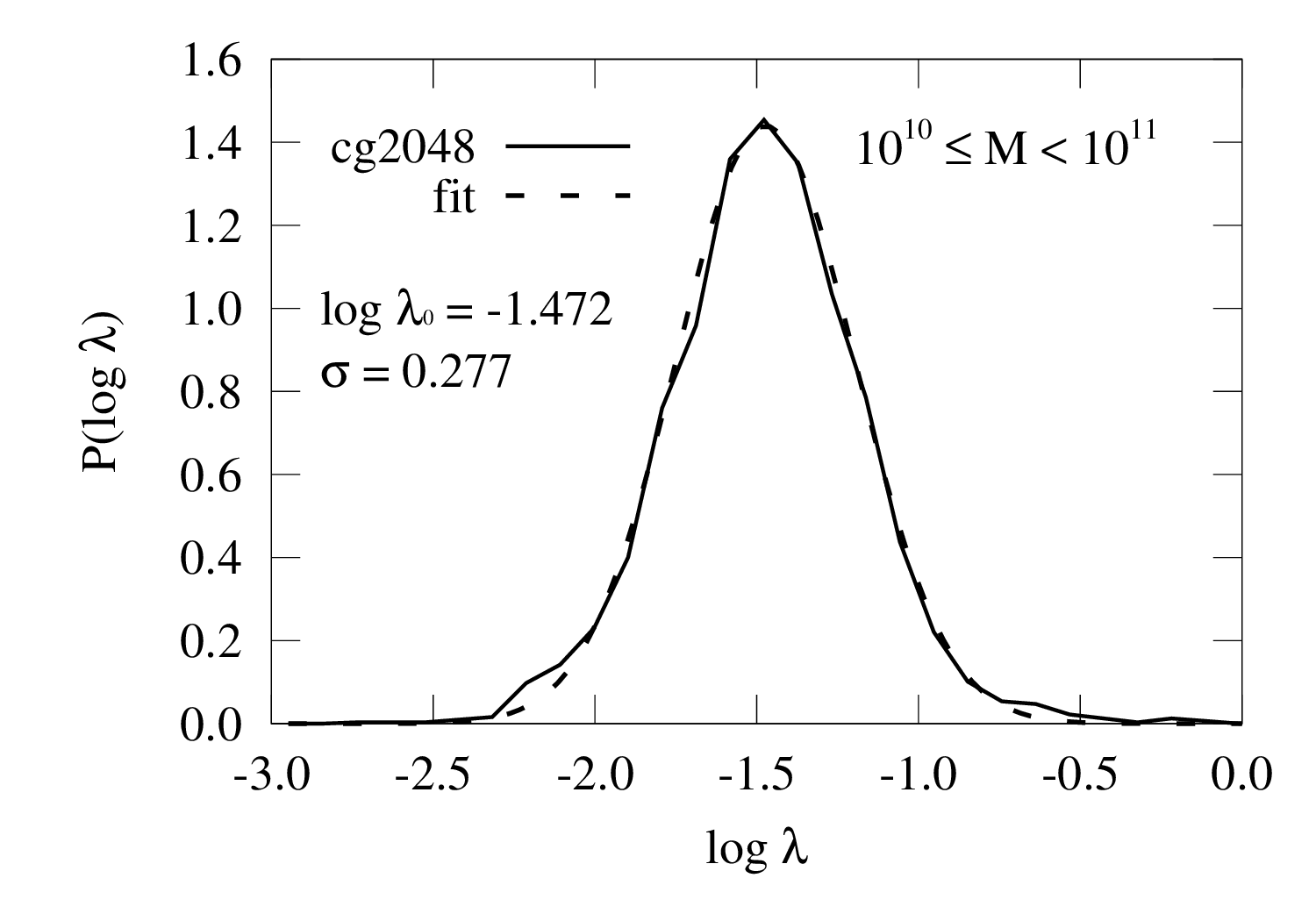}
\caption{
Probability distribution functions of the spin parameter at $z=0$.
These panels show the results of different mass ranges.
Dashed curves are the best fits of the log-normal distribution.
}
\label{fig:spin_pba}
\end{figure*}

First, we determine the minimum number of particles in a halo 
necessary to reliably determine the spin as done for the concentration. 
Figure \ref{fig:m-spin_res} shows the normalized difference 
of average spin between the CG2048 run and the CG512 run as a function of halo mass.
We can see that the difference is $\sim 0.05$ for halo mass 
larger than $8.0 \times 10^9 M_{\odot}$. 
For halo mass less than $8.0 \times 10^9 M_{\odot}$, the difference is large.
 
In the CG512 run, a halo of mass  $8.0 \times 10^9 M_{\odot}$ contains $\sim 1000$ particles.
So we conclude that we need $\sim 1000$ particles to reliably determine the concentration.
For the CG2048 run, the reliability limit is $1.28 \times 10^8M_{\odot}$.

Figure \ref{fig:m-spin} shows the median, and first third quantiles of
the spin parameter as a function of the virial mass of the halo.
Apparently, we can see the spin parameter is independent of
the mass down to $10^8 M_{\odot}$ as pointed out for larger halos in
previous works \citep{Maccio2007, Munoz2010}.  The median value is
0.0336.

Figure \ref{fig:spin_pba} shows the probability distribution functions 
of the spin parameter at $z=0$ in three different mass ranges. 
The distributions are well fitted by the log-normal distributions, 
\begin{eqnarray}
P(\log \lambda) = \frac{1}{\sqrt{2\pi} \sigma} 
\exp \left(-\frac{\log^2{(\lambda/\lambda_0)}}{2\sigma^2} \right).
\end{eqnarray}
We find $\log \lambda_0 = -1.477, \sigma = 0.308$
for halos with the mass of $1.28 \times 10^8 M_{\odot} \le M < 10^9 M_{\odot}$, 
$\log \lambda_0 = -1.480, \sigma = 0.288$
for halos with the mass of $10^9 M_{\odot} \le M < 10^{10} M_{\odot}$, and
$\log \lambda_0 = -1.472, \sigma = 0.277$
for halos with the mass of $10^{10} M_{\odot} \le M < 10^{11} M_{\odot}$.
Thus, we conclude that there is no mass dependence of the spin parameter. 
Otherwise, it is extremely weak.

We can see that there are small deviations from the log-normal distributions 
at high spin regions as seen in previous works for 
larger halos \citep{Bett2007, Delogu2010}.
We will discuss the effect of the dynamical state of halos in the Appendix.

\subsection{Subhalo}

\begin{figure}
\centering \includegraphics[width=8cm]{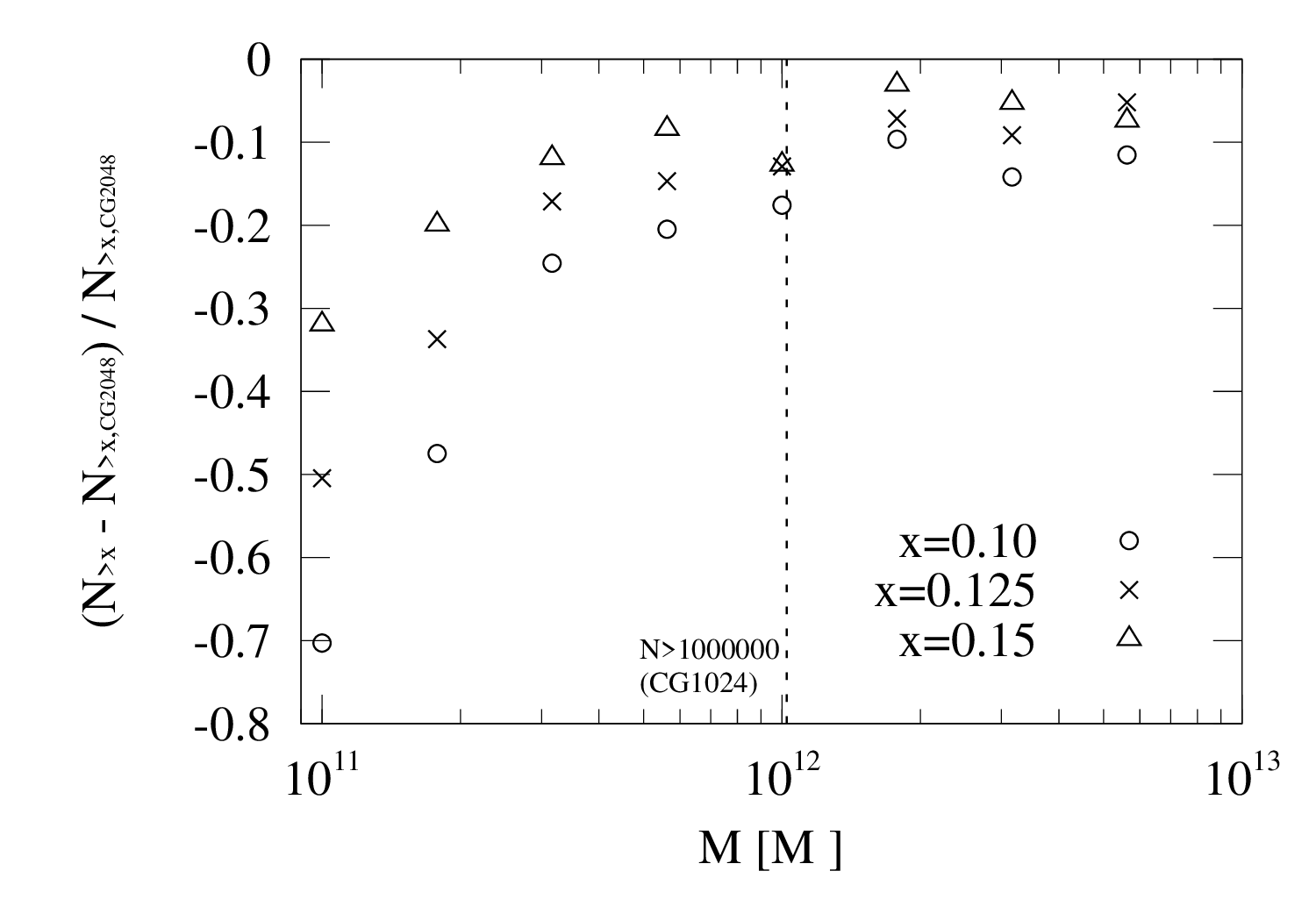} 
\caption{
Residuals of the subhalo abundance from the largest simulation (CG2048) to lower
resolution simulation (CG1024).  
Here, $N_{>x}$ is the number of subhalos with rotation velocity
larger than $x$ of that of the parent halo.
} 
\label{fig:m-nsub_res}
\end{figure}

\begin{figure}
\centering
\includegraphics[width=8cm]{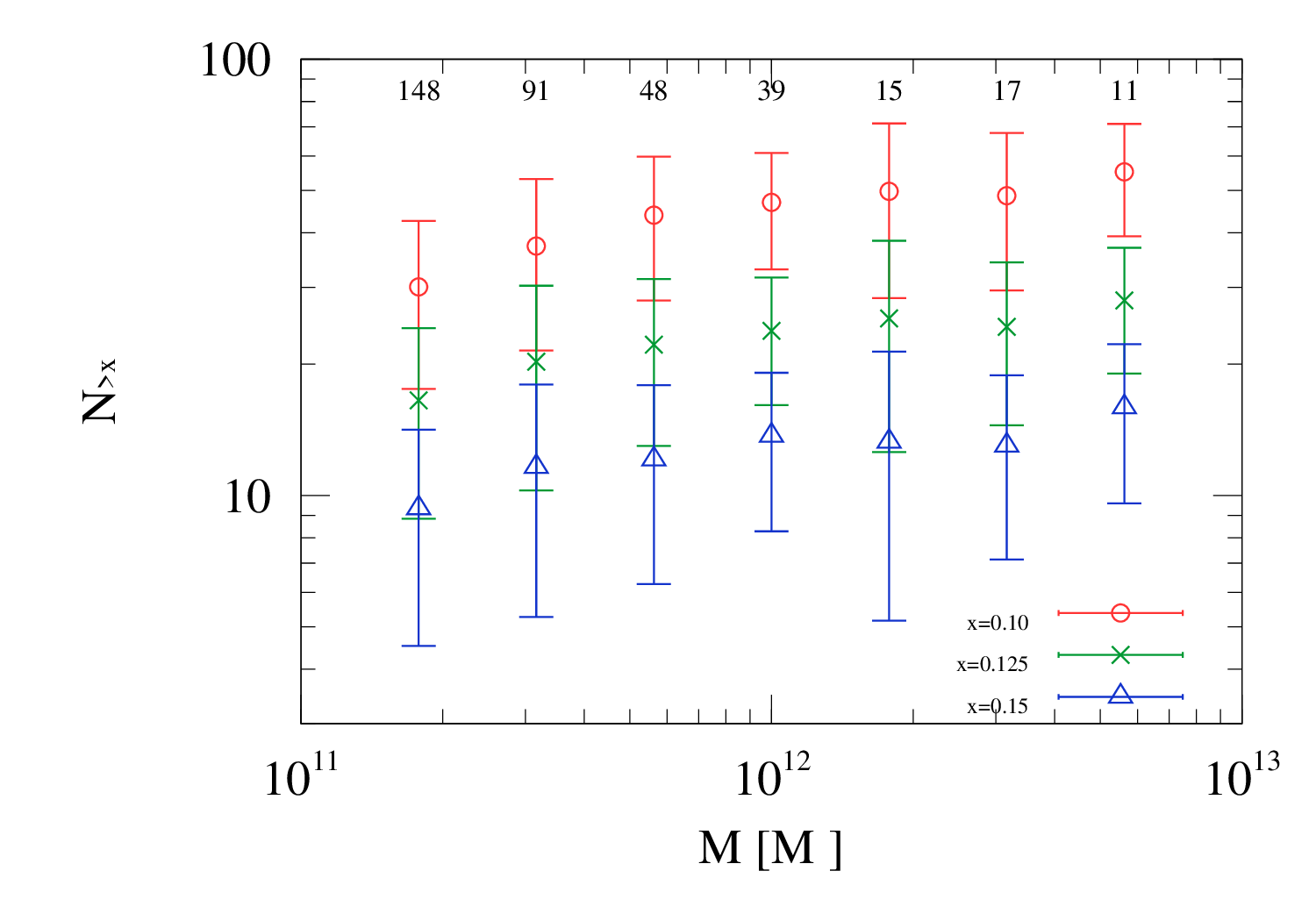}
\caption{
Number of subhalos plotted against the halo virial mass $M$ at $z=0$. 
Each symbol shows the mean value on each bin. 
Whiskers are the standard deviation.
The number of halos on each bin is shown above.
}
\label{fig:m-nsub}
\end{figure}

The statistics of the subhalo abundance of galaxy-sized halos have been
well studied \citep[e.g.,][]{Ishiyama2009,Boylan2010,Busha2011}.  The
subhalo abundance shows large halo-to-halo variations and depends on
the concentration parameter.  Halos with larger concentrations have
a smaller number of subhalos.  This means that the number of subhalos
should increase as the halo mass increases since the concentration
decreases.  However, little is known on 
how the subhalo abundance depends on the halo mass. 
The reason is that we need a number of
well-resolved halos in a wide mass range to determine the mass
dependence and it is computationally expensive to perform simulations
for this purpose.

\citet{Contini2012} analyzed the fraction of halo mass in subhalos for
group-sized to cluster-sized halos and showed that the fraction
increases with increasing mass.  
For group-sized halos, it is approximately 5\%, and for cluster-sized
halos approximately 10\% (similar results are obtained in
  \citep{Gao2011} for a slightly different mass range). However, the
number of particles per halo of their group-sized halos is $10^5$,
which is insufficient to robustly estimate the subhalo abundance (see
  also \citep{Ishiyama2009}).  Therefore, it is possible that they
have underestimated the subhalo abundance.

Our high-resolution simulations are suitable 
for the study of the statistics of 
the subhalo abundance for halos with smaller mass.
Therefore we can address
a key question, how the subhalo abundance depends on the halo mass.
Hereafter, we define $N_{>x}$ as the subhalo abundance, which is the
number of subhalos with rotation velocity larger than $x$ times that of
the parent halo.  Figure \ref{fig:m-nsub_res} shows the normalized
difference of average subhalo abundance between the CG2048 run and the
CG1024 run as a function of halo mass for $x=0.10, 0.125, 0.15$.  We
can see that both results are well converged for halos with more than
one million particles for all values of $x$.  For halos with
less particles, the difference becomes larger as the halo mass
decreases.  Thus, we can conclude that we need about one million
particles to reliably determine the subhalo abundance. For the CG2048
run, the reliability limit is $1.28 \times 10^{11} M_{\odot}$ for $x=0.1$.

The reliability limit should be smaller for larger subhalos (larger
values of $x$) since they consist of more particles than smaller ones.
As seen in Figure \ref{fig:m-nsub_res}, the residual of the subhalo
abundance is systematically smaller for larger subhalos (larger $x$).
However, for simplicity, we use the same reliability limit for all
values of $x$.  Thus, our choice of the reliability limit is quite
conservative.

Figure \ref{fig:m-nsub} shows the mean and the standard deviation of
the subhalo abundance as a function of the virial mass of the halo.
We can see clearly that the subhalo abundance depends on the halo mass
for all values of $x$.  The average number of subhalos $N_{>0.10},
N_{>0.125}, N_{>0.15}$ are 30.1, 16.5, 9.3 for halos with the mass of
$\sim 2 \times 10^{11} M_{\odot}$ and 47.0, 23.8, 13.7 for halos with
the mass of $\sim 1 \times 10^{12} M_{\odot}$.  For halos with the
mass of larger than $\sim 1 \times 10^{12} M_{\odot}$, we can see that
the dependence becomes weaker and gradually approaches to a constant
value.

This trend has not been observed in previous works \citep{Gao2011,
  Contini2012}, since they analyzed halos with larger mass.  
However, our result is limited by the box size of the simulation.
Since the number of subhalos with the mass larger than $1 \times
10^{12} M_{\odot}$ in our simulation is only 82, our halos within this
mass range might be a biased sample.  In order to clarify the
dependency, larger box simulations are needed.

\section{Discussions and Summary}\label{sec:discussion}

We present the first scientific results of the Cosmogrid simulation.  
Because of unprecedentedly high-resolution and
powerful statistics, the simulation is suitable to resolve internal
properties of halos with the mass larger than dwarf galaxy and subhalos
whose scales are comparable to ultra-faint dwarf galaxies.

We summarize the main results of this paper as follows.
\begin{itemize}
\item
The halo mass function is well described by the 
\citet{Sheth1999} fitting function down to 
$\sim 10^7 M_{\odot}$ from $1.0 \times 10^{13} M_{\odot}$. 
The differences are less than 10\% at $z=0$ from 
$M=5.0 \times 10^7 M_{\odot}$ to $M=2.0 \times 10^{12} M_{\odot}$. 

\item We analyzed the spherically averaged density profiles of the three
most massive halos which contain more than 170 million particles.  Their
masses are 5.24, 3.58, and 2.25 $\times 10^{13} M_{\odot}$.  We confirmed
that the slopes of density profiles of these halos become shallower than $-1$ at
the inner most radius. The results are consistent with the recent studies
based on high-resolution simulations for galactic halos.

\item We studied internal properties of halos at $z=0$ with the mass
more than $\sim 10^8 M_{\odot}$.  The concentration parameter measured by the
maximum rotational velocity radius is weakly correlated with the halo mass. 
We found that the dependence of the concentration parameter with halo mass 
cannot be expressed by a single power law, but levels off at small mass.
The slope of the mass$-$concentration relation is around $-0.07$ for 
halos with the mass $10^{10} M_{\odot}$, 
and $-0.06$ for halos with the mass $10^{9} M_{\odot}$.
The shallowing slope naturally emerges from 
the nature of the power spectrum of initial density fluctuations.
A simple model based on the Press$-$Schechter theory 
gives reasonable agreement with the simulation result.
The spin parameter does not show a correlation with the halo mass.  
The probability distribution functions of concentration and spin are well fitted by the
log-normal distribution for halos with the mass 
larger than $\sim 10^8 M_{\odot}$.

\item
The subhalo abundance depends on the halo mass. 
Galaxy-sized halos have 50\% more subhalos than 
$\sim 10^{11} M_{\odot}$ halos have.
We find a new result that the dependence
becomes weaker for more massive halos.

\end{itemize}

\acknowledgements
We thank the anonymous referee for his/her valuable comments.
Numerical computations were partially carried out on Cray XT4 at Center
for Computational Astrophysics, CfCA, of National Astronomical
Observatory of Japan, 
the K computer at the RIKEN Advanced Institute for Computational Science 
(Proposal number hp120286), 
Huygens at the Dutch National High Performance
Computing and e-Science Support Center, SARA (Netherlands),  
HECToR at the Edinburgh Parallel Computing Center (United Kingdom), 
and Louhi at IT Center for Science in Espoo (Finland).
T.I. is financially supported by Research
Fellowship of the Japan Society for the Promotion of Science (JSPS) for
Young Scientists.  This research is partially supported by the Special
Coordination Fund for Promoting Science and Technology (GRAPE-DR
project), Ministry of Education, Culture, Sports, Science and
Technology, Japan.  We also thank the network facilities of SURFnet,
DEISA, IEEAF, WIDE, Northwest Gigapop and the Global Lambda Integrated
FAcility (GLIF) GOLE of TransLight Cisco on National LambdaRail, Trans-
Light, StarLight, NetherLight, T-LEX, Pacific and Atlantic Wave. This
research is supported by the Netherlands organization for Scientific
research (NWO) grant Nos.639.073.803, 643.200.503 and
643.000.803, the Stichting Nationale Computerfaciliteiten
(project SH-095-08), NAOJ, SURFnet (GigaPort project), the
International Information Science Foundation (IISF), the Netherlands
Advanced School for Astronomy (NOVA), the Leids Kerkhoven-Bosscha fonds
(LKBF). We thank the DEISA Consortium (EU FP6 project RI-031513 and FP7
project RI-222919) for support within the DEISA Extreme Computing
Initiative (GBBP project).
This work has been funded by MEXT HPCI STRATEGIC PROGRAM 
and MEXT/JSPS KAKENHI Grand Number 24740115.

\appendix
\section{The Effect of Dynamical State of Halos}
There are large variations in the dynamical state of halos.  Halos
which formed in an early epoch tend to be dynamically relaxed, whereas
halos which experienced a recent major merger tend to be dynamically
unrelaxed. The relaxation state of halos might have some effect on
properties of halos such as the concentration and the spin.  Here, we
analyze these properties for only dynamically relaxed sample of halos
and discuss the effect of the relaxation state.

\citet{Power2012} argued that the center-of-mass offset is a robust estimator 
of the relaxation state of halos. The center-of-mass offset is defined as 
\begin{equation}
\Delta r = \frac{|{\bm r_{\rm cen}} - {\bm r_{\rm cm}}|}{R_{\rm vir}},
\end{equation}
where ${\bm r_{\rm cen}}$, ${\bm r_{\rm cm}}$, and $R_{\rm vir}$ are
the center of density, mass, and the virial radius of a halo.  They
found that $\Delta r \le 0.04$ is a sufficient condition to pick up
dynamically relaxed halos at $z=0$.  We use this condition to
construct the relaxed sample of halos from our all halo samples.

\begin{figure}
\centering
\includegraphics[width=6.7cm]{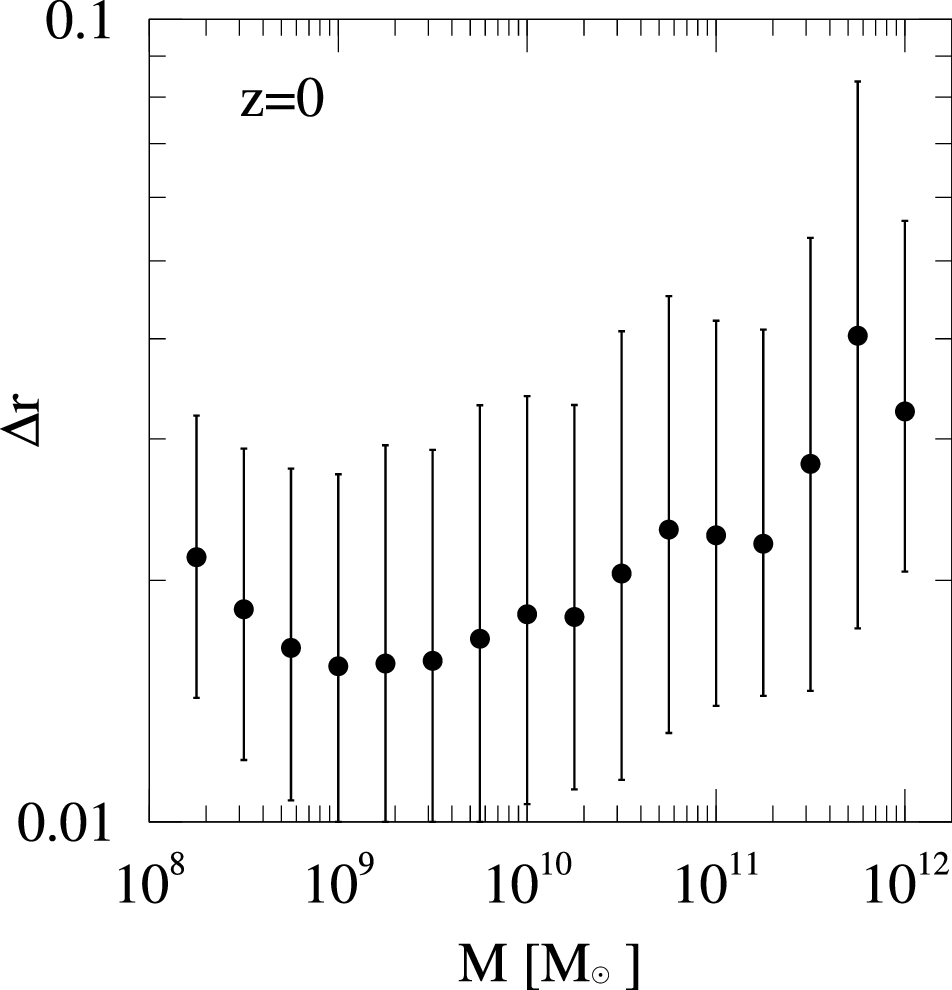}
\includegraphics[width=7cm]{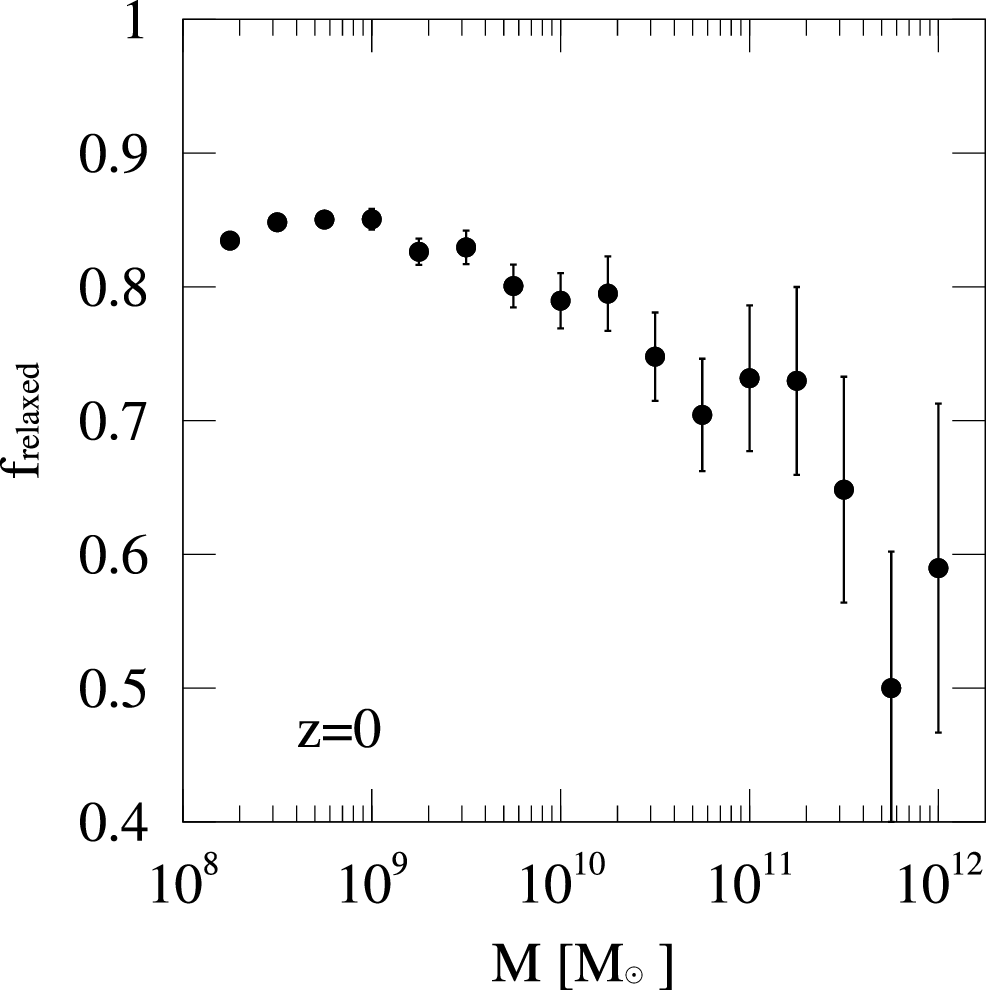}
\caption{
Left: the center-of-mass offset of all halos
plotted against  the halo virial mass $M$ at $z=0$.
The median value of each bin is shown by circles. 
Whiskers are the first and third quantiles. 
Right: the fraction of relaxed halos.
Error bars are Poisson errors.
}
\label{fig:relaxed_num}
\end{figure}

Figure \ref{fig:relaxed_num} shows the average center-of-mass offset
and the fraction of relaxed halos as a function of the halo virial
mass. The offset increases with increasing the halo mass.  This trend
is in good agreement with the results of \cite{Power2012}.  It is
simply because lower mass halos tend to form earlier than higher mass
halos from the nature of the hierarchical structure formation.  As a
result, the fraction of relaxed halos becomes large for lower mass
halos. We can see the offset increases with decreasing the halo mass 
from $\sim 5 \times 10^8M_{\odot}$. 
This may be caused by the resolution effect. 

\begin{figure}
\centering
\includegraphics[width=8cm]{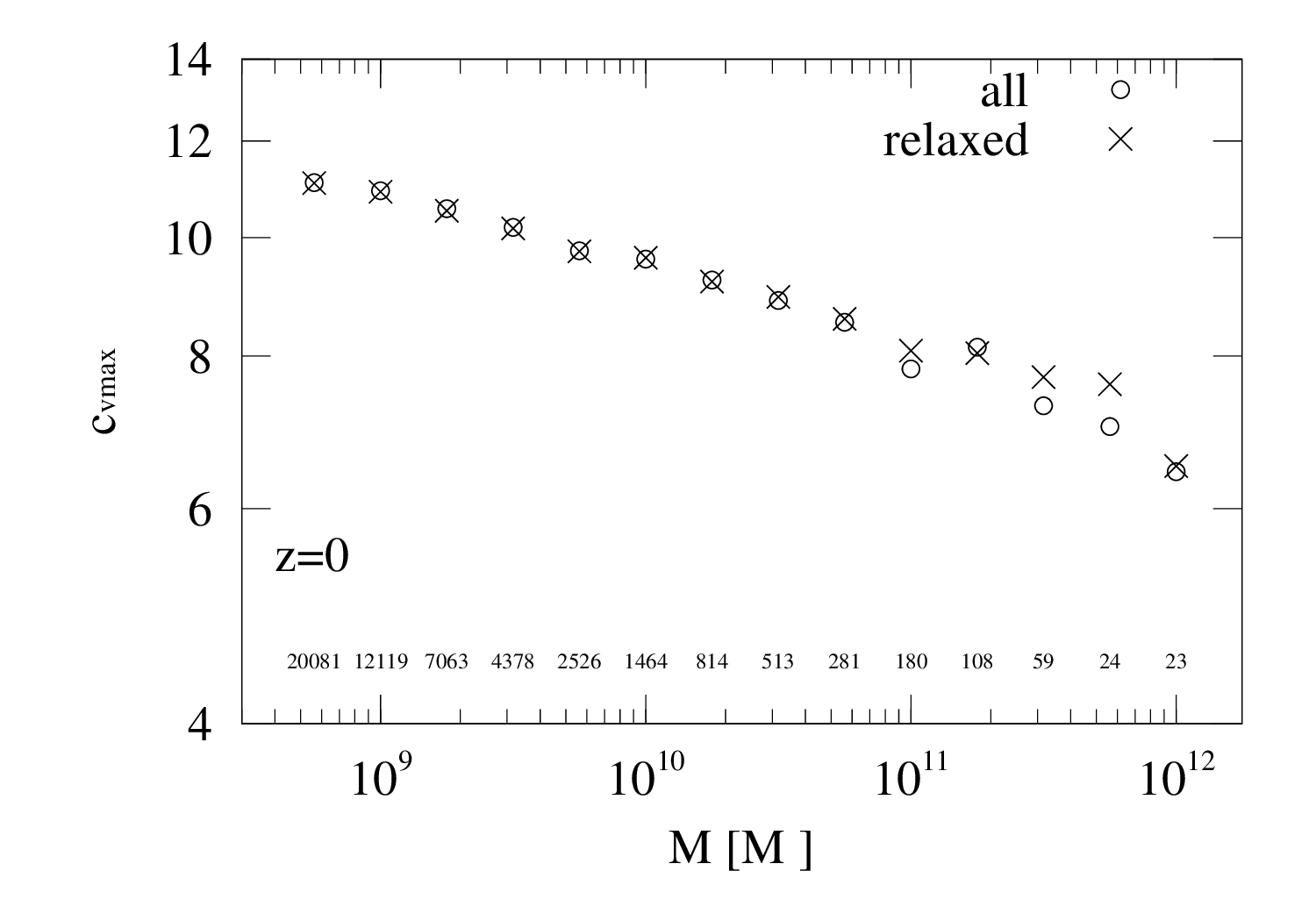}
\includegraphics[width=8cm]{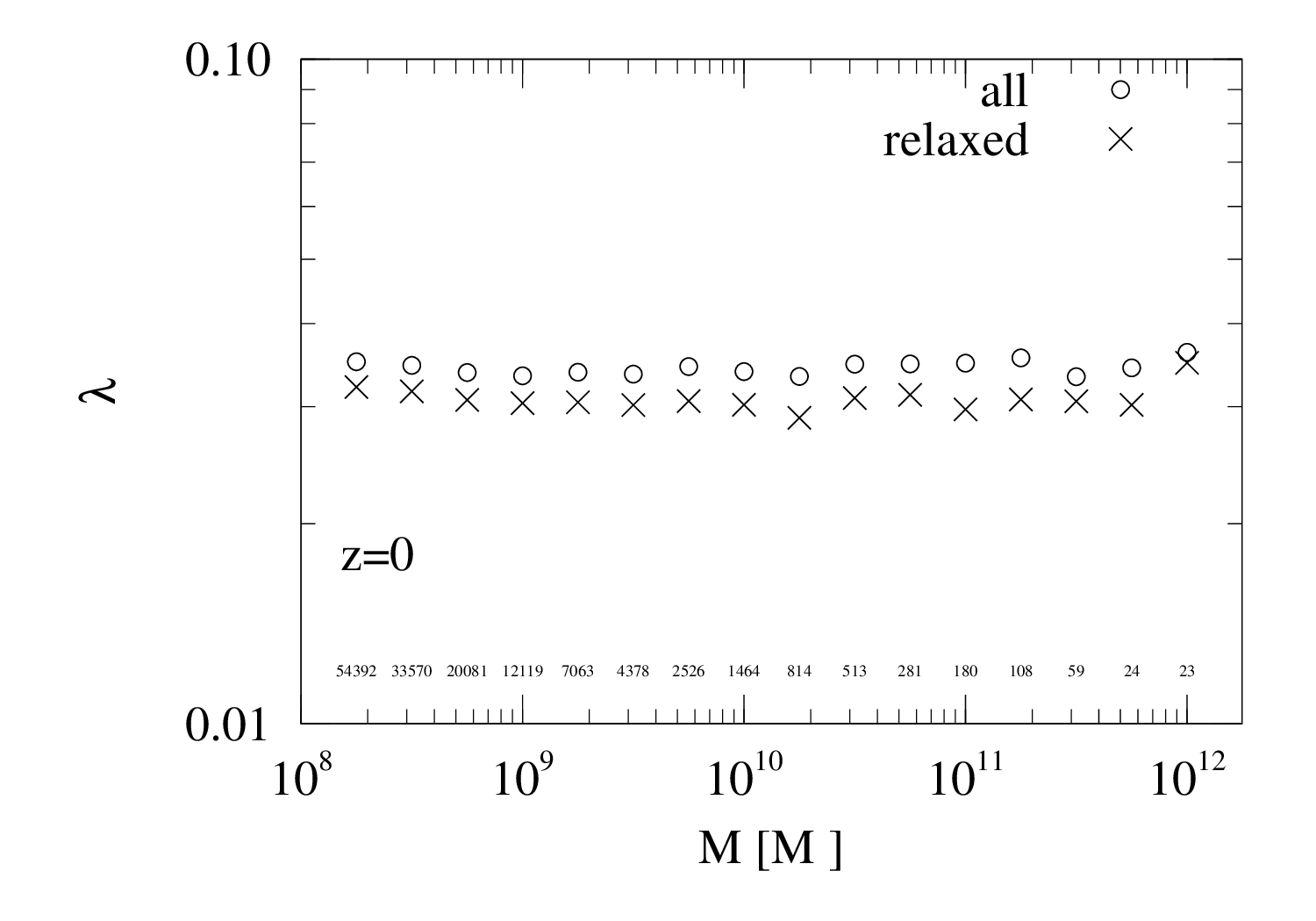}
\caption{
Concentration (left) and spin (right) plotted 
against the halo virial mass $M$ at $z=0$.
The median values of all halos are shown by circles. 
Crosses are for the values of only relaxed halos.
The number of relaxed halos on each bin is shown below.
}
\label{fig:relaxed}
\end{figure}

One may wonder whether the dependence of concentrations to the halo
mass is caused by unrelaxed halos or not. Figure \ref{fig:relaxed}
shows the median concentration and spin for all and the relaxed sample
of halos as a function of the virial mass of the halo. The relaxation
state has little impact on the concentration for halos smaller than
$10^{11}M_{\odot}$.  This can be interpreted as the fact that the
fraction of relaxed halos is large for lower mass halos as we can see
in Figure \ref{fig:relaxed_num}.

\begin{figure}
\centering
\includegraphics[width=5.9cm]{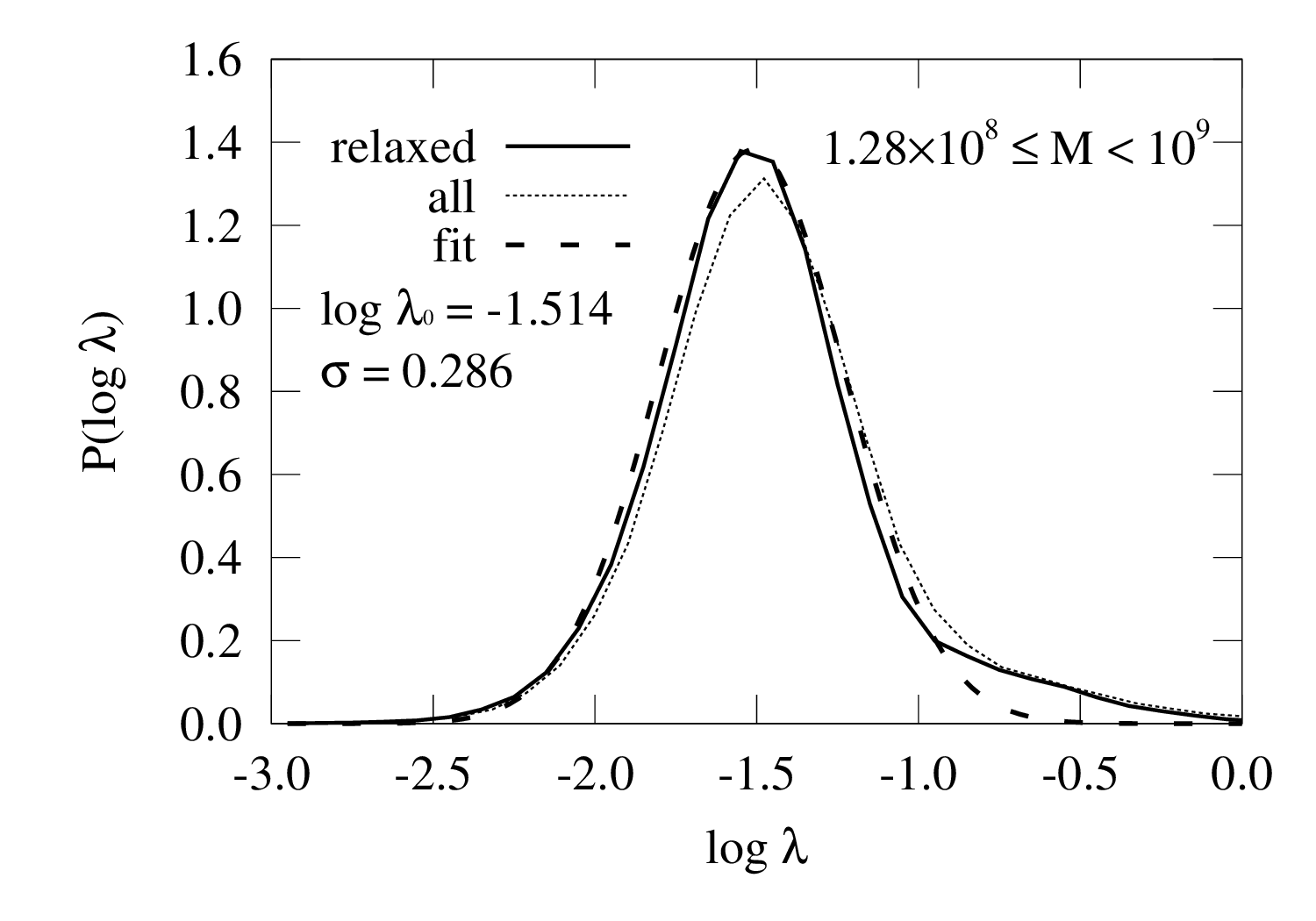}
\includegraphics[width=5.9cm]{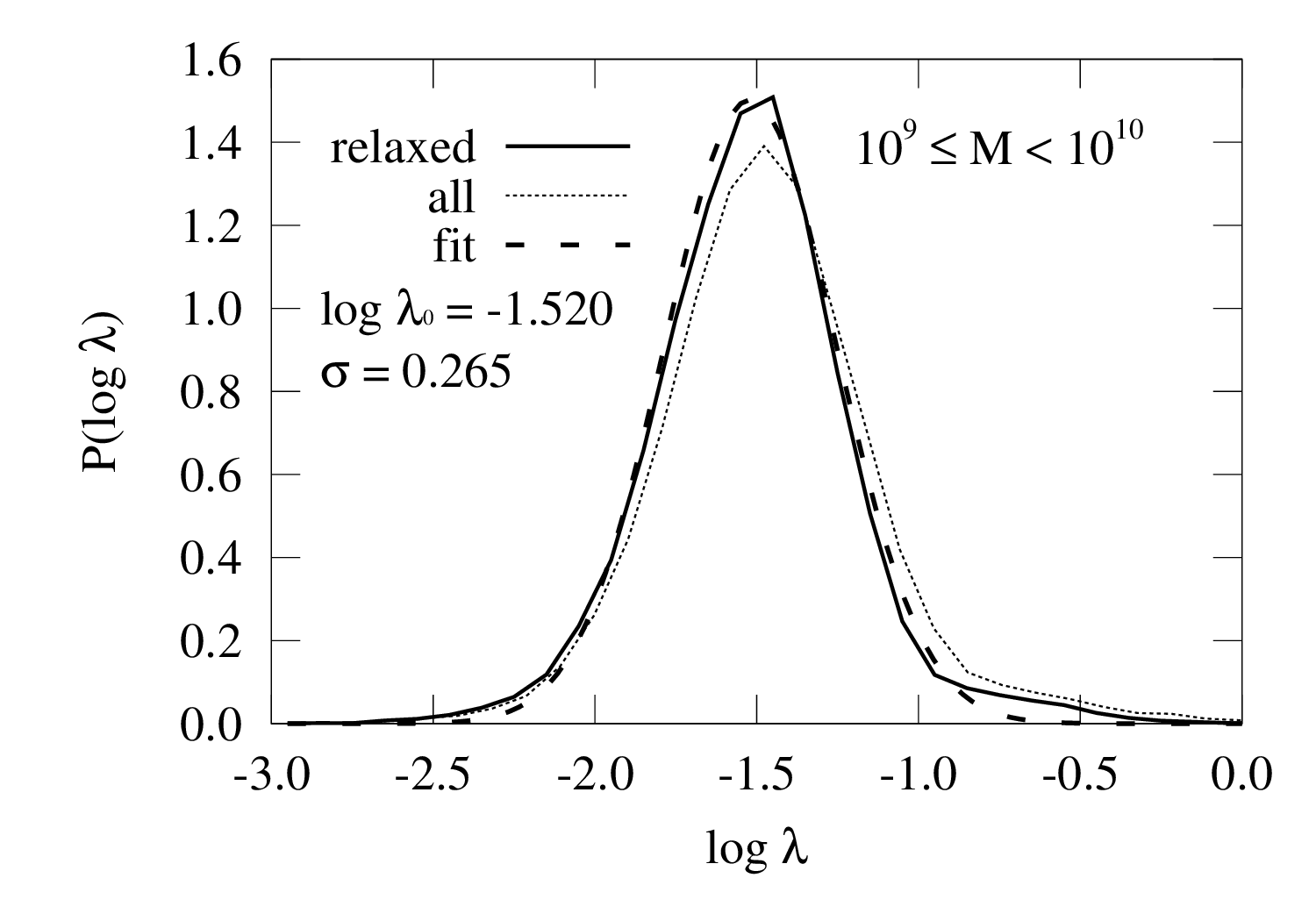}
\includegraphics[width=5.9cm]{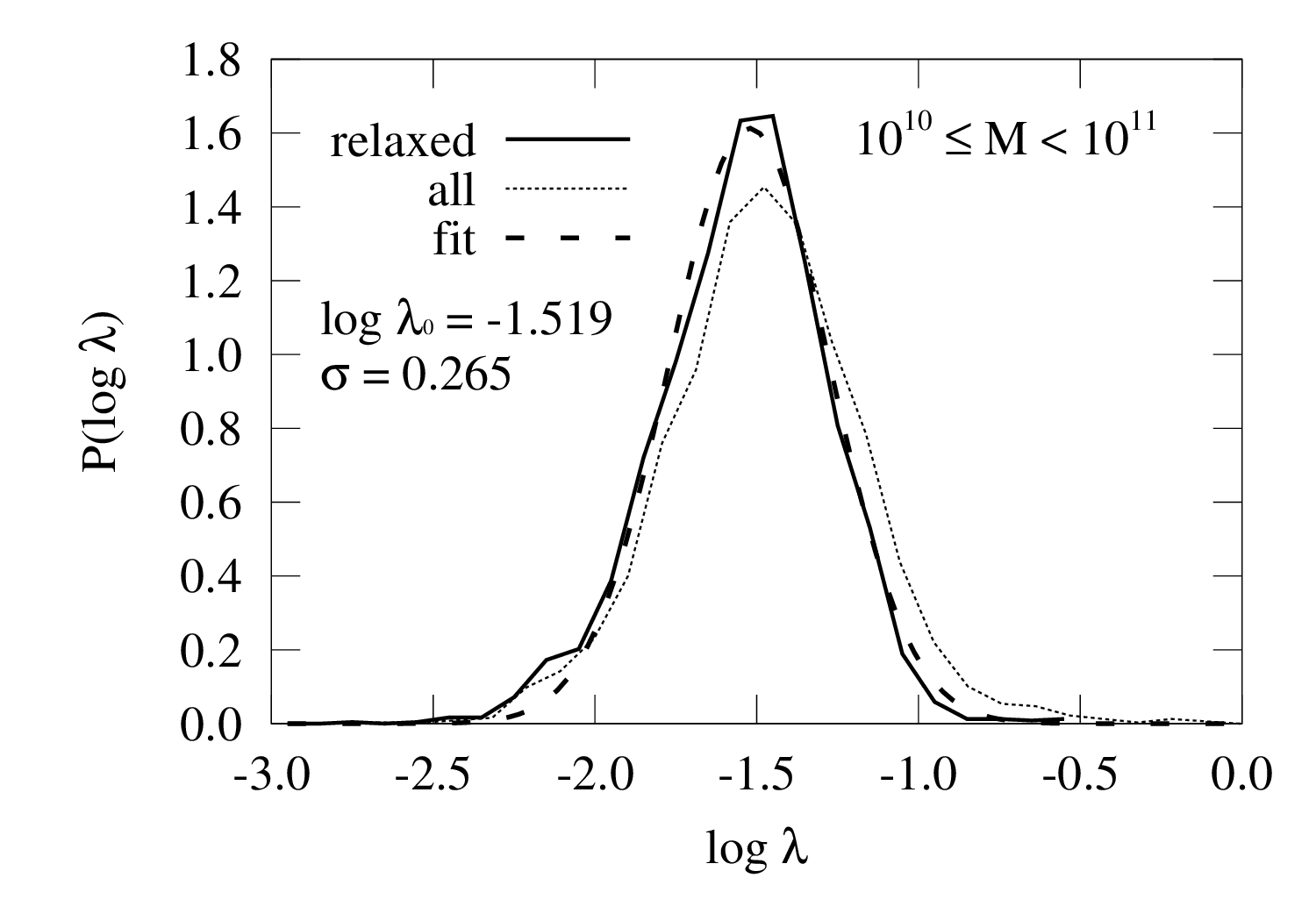}
\caption{
Probability distribution functions of the spin parameter at $z=0$
for relaxed halos (solid) and all halos (dotted).
Dashed curves are the best fits of the log-normal distribution
for relaxed halos.
}
\label{fig:spin_relaxed}
\end{figure}

The spin parameters of relaxed halos are systematically smaller than
those of all halos by $\sim 8-10 \%$ for all mass ranges.  This result
is consistent with early studies \citep{Maccio2007,Maccio2008}.  
This is because unrelaxed halos tend to experience a recent major merger, 
giving them higher spin values. 
Figure \ref{fig:spin_relaxed} shows the
probability distribution functions of the spin parameter at $z=0$ in
three different mass ranges.  We can see clearly that the number of
halos with high spin values in the relaxed sample of halos is smaller than
that in all sample of halos for all mass ranges.  We find $\log
\lambda_0 = -1.514, \sigma = 0.286$ for halos with the mass of $1.28
\times 10^8 M_{\odot} \le M < 10^9 M_{\odot}$, $\log \lambda_0 =
-1.520, \sigma = 0.265$ for halos with the mass of $10^9 M_{\odot} \le
M < 10^{10} M_{\odot}$, and $\log \lambda_0 = -1.519, \sigma = 0.265$
for halos with the mass of $10^{10} M_{\odot} \le M < 10^{11}
M_{\odot}$.  The standard deviations are also systematically smaller
for relaxed halos by $\sim 4-8 \%$. 

Small deviations from the log-normal distributions at high spin
regions for all halos are also seen for relaxed halos. 
The deviations become weaker since unrelaxed halos with
higher spin are removed. 

It is interesting that the spin is relatively influenced by the
relaxation state more than the concentration.  This might be because
halos grow in a self-similar way \citep[e.g.,][]{Fukushige2001}.  The
self-similar growth means that the inner region of a halo forms
earlier than the outer region.  Here, the spin is calculated using all
particles.  The concentration is estimated using particles within the
radius of the maximum rotational velocity, which should be more
dynamically relaxed than particles in outer region.  Therefore, it is
natural that the effect of the relaxation state on the concentration
and spin shows such difference.

In summary, we find that the relaxation state makes small difference 
on the concentration and spin distributions. 
\begin{itemize}
\item
The impact of the relaxation state on the concentration is negligible 
for halos smaller than $10^{11}M_{\odot}$. 
\item
The spin parameters of relaxed halos are systematically smaller than
those of all halos by $\sim 8-10 \%$ for all mass ranges.  
The spin distributions of relaxed halos deviate from the log-normal fitting
 less than those of all halos.
\end{itemize}

\begin{thebibliography}

\bibitem[{{Aarseth} {et~al.}(1979){Aarseth}, {Turner}, \& {Gott}}]{Aarseth1979}
{Aarseth}, S.~J., {Turner}, E.~L., \& {Gott}, III, J.~R. 1979, \apj, 228, 664

\bibitem[{{Antonuccio-Delogu} {et~al.}(2010){Antonuccio-Delogu}, {Dobrotka},
  {Becciani}, {Cielo}, {Giocoli}, {Macci{\`o}}, \&
  {Romeo-Velon{\'a}}}]{Delogu2010}
{Antonuccio-Delogu}, V., {Dobrotka}, A., {Becciani}, U., {Cielo}, S.,
  {Giocoli}, C., {Macci{\`o}}, A.~V., \& {Romeo-Velon{\'a}}, A. 2010, \mnras,
  1016

\bibitem[{{Bagla} \& {Prasad}(2006)}]{Bagla2006}
{Bagla}, J.~S., \& {Prasad}, J. 2006, \mnras, 370, 993

\bibitem[{{Bailin} \& {Steinmetz}(2005)}]{Bailin2005}
{Bailin}, J., \& {Steinmetz}, M. 2005, \apj, 627, 647

\bibitem[{{Barnes} \& {Hut}(1986)}]{Barnes1986}
{Barnes}, J., \& {Hut}, P. 1986, \nat, 324, 446

\bibitem[{{Bertschinger}(2001)}]{Bertschinger2001}
{Bertschinger}, E. 2001, \apjs, 137, 1

\bibitem[{{Bett} {et~al.}(2007){Bett}, {Eke}, {Frenk}, {Jenkins}, {Helly}, \&
  {Navarro}}]{Bett2007}
{Bett}, P., {Eke}, V., {Frenk}, C.~S., {Jenkins}, A., {Helly}, J., \&
  {Navarro}, J. 2007, \mnras, 376, 215

\bibitem[{{Boylan-Kolchin} {et~al.}(2010){Boylan-Kolchin}, {Springel}, {White},
  \& {Jenkins}}]{Boylan2010}
{Boylan-Kolchin}, M., {Springel}, V., {White}, S.~D.~M., \& {Jenkins}, A. 2010,
  \mnras, 406, 896

\bibitem[{{Boylan-Kolchin} {et~al.}(2009){Boylan-Kolchin}, {Springel}, {White},
  {Jenkins}, \& {Lemson}}]{Boylan2009}
{Boylan-Kolchin}, M., {Springel}, V., {White}, S.~D.~M., {Jenkins}, A., \&
  {Lemson}, G. 2009, \mnras, 398, 1150

\bibitem[{{Bryan} \& {Norman}(1998)}]{Bryan1998}
{Bryan}, G.~L., \& {Norman}, M.~L. 1998, \apj, 495, 80

\bibitem[{{Bullock} {et~al.}(2001{\natexlab{a}}){Bullock}, {Dekel}, {Kolatt},
  {Kravtsov}, {Klypin}, {Porciani}, \& {Primack}}]{Bullock2001b}
{Bullock}, J.~S., {Dekel}, A., {Kolatt}, T.~S., {Kravtsov}, A.~V., {Klypin},
  A.~A., {Porciani}, C., \& {Primack}, J.~R. 2001{\natexlab{a}}, \apj, 555, 240

\bibitem[{{Bullock} {et~al.}(2001{\natexlab{b}}){Bullock}, {Kolatt}, {Sigad},
  {Somerville}, {Kravtsov}, {Klypin}, {Primack}, \& {Dekel}}]{Bullock2001}
{Bullock}, J.~S., {Kolatt}, T.~S., {Sigad}, Y., {Somerville}, R.~S.,
  {Kravtsov}, A.~V., {Klypin}, A.~A., {Primack}, J.~R., \& {Dekel}, A.
  2001{\natexlab{b}}, \mnras, 321, 559

\bibitem[{{Busha} {et~al.}(2011){Busha}, {Wechsler}, {Behroozi}, {Gerke},
  {Klypin}, \& {Primack}}]{Busha2011}
{Busha}, M.~T., {Wechsler}, R.~H., {Behroozi}, P.~S., {Gerke}, B.~F., {Klypin},
  A.~A., \& {Primack}, J.~R. 2011, \apj, 743, 117

\bibitem[{{Contini} {et~al.}(2012){Contini}, {De Lucia}, \&
  {Borgani}}]{Contini2012}
{Contini}, E., {De Lucia}, G., \& {Borgani}, S. 2012, \mnras, 420, 2978

\bibitem[{{Crocce} {et~al.}(2010){Crocce}, {Fosalba}, {Castander}, \&
  {Gazta{\~n}aga}}]{Crocce2010}
{Crocce}, M., {Fosalba}, P., {Castander}, F.~J., \& {Gazta{\~n}aga}, E. 2010,
  \mnras, 403, 1353

\bibitem[{{Diemand} {et~al.}(2007){Diemand}, {Kuhlen}, \&
  {Madau}}]{Diemand2007}
{Diemand}, J., {Kuhlen}, M., \& {Madau}, P. 2007, \apj, 657, 262

\bibitem[{{Diemand} {et~al.}(2008){Diemand}, {Kuhlen}, {Madau}, {Zemp},
  {Moore}, {Potter}, \& {Stadel}}]{Diemand2008}
{Diemand}, J., {Kuhlen}, M., {Madau}, P., {Zemp}, M., {Moore}, B., {Potter},
  D., \& {Stadel}, J. 2008, \nat, 454, 735

\bibitem[{{Diemand} {et~al.}(2004){Diemand}, {Moore}, \&
  {Stadel}}]{Diemand2004b}
{Diemand}, J., {Moore}, B., \& {Stadel}, J. 2004, \mnras, 353, 624

\bibitem[{{Diemand} {et~al.}(2005){Diemand}, {Zemp}, {Moore}, {Stadel}, \&
  {Carollo}}]{Diemand2005b}
{Diemand}, J., {Zemp}, M., {Moore}, B., {Stadel}, J., \& {Carollo}, C.~M. 2005,
  \mnras, 364, 665

\bibitem[{{Efstathiou}(1979)}]{Efstathiou1979}
{Efstathiou}, G. 1979, \mnras, 187, 117

\bibitem[{{Evrard} {et~al.}(2002){Evrard}, {MacFarland}, {Couchman}, {Colberg},
  {Yoshida}, {White}, {Jenkins}, {Frenk}, {Pearce}, {Peacock}, \&
  {Thomas}}]{Evrard2002}
{Evrard}, A.~E., {et~al.} 2002, \apj, 573, 7

\bibitem[{{Fall}(1978)}]{Fall1978}
{Fall}, S.~M. 1978, \mnras, 185, 165

\bibitem[{{Fukushige} {et~al.}(2004){Fukushige}, {Kawai}, \&
  {Makino}}]{Fukushige2004}
{Fukushige}, T., {Kawai}, A., \& {Makino}, J. 2004, \apj, 606, 625

\bibitem[{{Fukushige} \& {Makino}(1997)}]{Fukushige1997}
{Fukushige}, T., \& {Makino}, J. 1997, \apjl, 477, L9+

\bibitem[{{Fukushige} \& {Makino}(2001)}]{Fukushige2001}
{Fukushige}, T., \& {Makino}, J. 2001, \apj, 557, 533

\bibitem[{{Fukushige} \& {Makino}(2003)}]{Fukushige2003}
{Fukushige}, T., \& {Makino}, J. 2003, \apj, 588, 674

\bibitem[{{Gao} {et~al.}(2011){Gao}, {Frenk}, {Boylan-Kolchin}, {Jenkins},
  {Springel}, \& {White}}]{Gao2011}
{Gao}, L., {Frenk}, C.~S., {Boylan-Kolchin}, M., {Jenkins}, A., {Springel}, V.,
  \& {White}, S.~D.~M. 2011, \mnras, 410, 2309

\bibitem[{{Gao} {et~al.}(2008){Gao}, {Navarro}, {Cole}, {Frenk}, {White},
  {Springel}, {Jenkins}, \& {Neto}}]{Gao2008}
{Gao}, L., {Navarro}, J.~F., {Cole}, S., {Frenk}, C.~S., {White}, S.~D.~M.,
  {Springel}, V., {Jenkins}, A., \& {Neto}, A.~F. 2008, \mnras, 387, 536

\bibitem[{{Gao} {et~al.}(2004){Gao}, {White}, {Jenkins}, {Stoehr}, \&
  {Springel}}]{Gao2004}
{Gao}, L., {White}, S.~D.~M., {Jenkins}, A., {Stoehr}, F., \& {Springel}, V.
  2004, \mnras, 355, 819

\bibitem[{{Ghigna} {et~al.}(2000){Ghigna}, {Moore}, {Governato}, {Lake},
  {Quinn}, \& {Stadel}}]{Ghigna2000}
{Ghigna}, S., {Moore}, B., {Governato}, F., {Lake}, G., {Quinn}, T., \&
  {Stadel}, J. 2000, \apj, 544, 616

\bibitem[{{Graham} {et~al.}(2006){Graham}, {Merritt}, {Moore}, {Diemand}, \&
  {Terzi{\'c}}}]{Graham2006}
{Graham}, A.~W., {Merritt}, D., {Moore}, B., {Diemand}, J., \& {Terzi{\'c}}, B.
  2006, \aj, 132, 2701

\bibitem[{{Groen} {et~al.}(2011){Groen}, {Portegies Zwart}, {Ishiyama}, \&
  {Makino}}]{Groen2011}
{Groen}, D., {Portegies Zwart}, S., {Ishiyama}, T., \& {Makino}, J. 2011,
  Computational Science and Discovery, 4, 015001

\bibitem[{{Groen} {et~al.}(2010){Groen}, {Rieder}, {Grosso}, {de Laat}, \&
  {Portegies Zwart}}]{Groen2010}
{Groen}, D., {Rieder}, S., {Grosso}, P., {de Laat}, C., \& {Portegies Zwart},
  S. 2010, Computational Science and Discovery, 3, 015002

\bibitem[{{Hayashi} {et~al.}(2004){Hayashi}, {Navarro}, {Power}, {Jenkins},
  {Frenk}, {White}, {Springel}, {Stadel}, \& {Quinn}}]{Hayashi2004}
{Hayashi}, E., {et~al.} 2004, \mnras, 355, 794

\bibitem[{{Hockney} \& {Eastwood}(1981)}]{Hockney1981}
{Hockney}, R.~W., \& {Eastwood}, J.~W. 1981, {Computer Simulation Using
  Particles (New York: McGraw-Hill)}

\bibitem[{{Ishiyama} {et~al.}(2008){Ishiyama}, {Fukushige}, \&
  {Makino}}]{Ishiyama2008}
{Ishiyama}, T., {Fukushige}, T., \& {Makino}, J. 2008, \pasj, 60, L13+

\bibitem[{{Ishiyama} {et~al.}(2009{\natexlab{a}}){Ishiyama}, {Fukushige}, \&
  {Makino}}]{Ishiyama2009b}
{Ishiyama}, T., {Fukushige}, T., \& {Makino}, J. 2009{\natexlab{a}}, \pasj, 61,
  1319

\bibitem[{{Ishiyama} {et~al.}(2009{\natexlab{b}}){Ishiyama}, {Fukushige}, \&
  {Makino}}]{Ishiyama2009}
{Ishiyama}, T., {Fukushige}, T., \& {Makino}, J. 2009{\natexlab{b}}, \apj, 696,
  2115

\bibitem[{{Jenkins} {et~al.}(2001){Jenkins}, {Frenk}, {White}, {Colberg},
  {Cole}, {Evrard}, {Couchman}, \& {Yoshida}}]{Jenkins2001}
{Jenkins}, A., {Frenk}, C.~S., {White}, S.~D.~M., {Colberg}, J.~M., {Cole}, S.,
  {Evrard}, A.~E., {Couchman}, H.~M.~P., \& {Yoshida}, N. 2001, \mnras, 321,
  372

\bibitem[{{Jing}(2000)}]{Jing2000b}
{Jing}, Y.~P. 2000, \apj, 535, 30

\bibitem[{{Jing} \& {Suto}(2000)}]{Jing2000a}
{Jing}, Y.~P., \& {Suto}, Y. 2000, \apjl, 529, L69

\bibitem[{{Jing} \& {Suto}(2002)}]{Jing2002}
{Jing}, Y.~P., \& {Suto}, Y. 2002, \apj, 574, 538

\bibitem[{{Kawai} {et~al.}(2004){Kawai}, {Makino}, \& {Ebisuzaki}}]{Kawai2004}
{Kawai}, A., {Makino}, J., \& {Ebisuzaki}, T. 2004, \apjs, 151, 13

\bibitem[{{Kazantzidis} {et~al.}(2006){Kazantzidis}, {Zentner}, \&
  {Kravtsov}}]{Kazantzidis2006}
{Kazantzidis}, S., {Zentner}, A.~R., \& {Kravtsov}, A.~V. 2006, \apj, 641, 647

\bibitem[{{Kim} {et~al.}(2009){Kim}, {Park}, {Gott}, \& {Dubinski}}]{Kim2009}
{Kim}, J., {Park}, C., {Gott}, J.~R., \& {Dubinski}, J. 2009, \apj, 701, 1547

\bibitem[{{Klypin} {et~al.}(2001){Klypin}, {Kravtsov}, {Bullock}, \&
  {Primack}}]{Klypin2001}
{Klypin}, A., {Kravtsov}, A.~V., {Bullock}, J.~S., \& {Primack}, J.~R. 2001,
  \apj, 554, 903

\bibitem[{{Klypin} {et~al.}(1999){Klypin}, {Kravtsov}, {Valenzuela}, \&
  {Prada}}]{Klypin1999}
{Klypin}, A., {Kravtsov}, A.~V., {Valenzuela}, O., \& {Prada}, F. 1999, \apj,
  522, 82

\bibitem[{{Klypin} {et~al.}(2011){Klypin}, {Trujillo-Gomez}, \&
  {Primack}}]{Klypin2011}
{Klypin}, A.~A., {Trujillo-Gomez}, S., \& {Primack}, J. 2011, \apj, 740, 102

\bibitem[{{Knebe} \& {Power}(2008)}]{Knebe2008}
{Knebe}, A., \& {Power}, C. 2008, \apj, 678, 621

\bibitem[{{Kroupa} {et~al.}(2010){Kroupa}, {Famaey}, {de Boer},
  {Dabringhausen}, {Pawlowski}, {Boily}, {Jerjen}, {Forbes}, {Hensler}, \&
  {Metz}}]{Kroupa2010}
{Kroupa}, P., {et~al.} 2010, \aap, 523, A32+

\bibitem[{{Lacey} \& {Cole}(1993)}]{Lacey1993}
{Lacey}, C., \& {Cole}, S. 1993, \mnras, 262, 627

\bibitem[{{Lacey} \& {Cole}(1994)}]{Lacey1994}
{Lacey}, C., \& {Cole}, S. 1994, \mnras, 271, 676

\bibitem[{{Li} {et~al.}(2009){Li}, {Helmi}, {De Lucia}, \& {Stoehr}}]{Li2009}
{Li}, Y., {Helmi}, A., {De Lucia}, G., \& {Stoehr}, F. 2009, \mnras, 397, L87

\bibitem[{{Luki{\'c}} {et~al.}(2007){Luki{\'c}}, {Heitmann}, {Habib},
  {Bashinsky}, \& {Ricker}}]{Lukic2007}
{Luki{\'c}}, Z., {Heitmann}, K., {Habib}, S., {Bashinsky}, S., \& {Ricker},
  P.~M. 2007, \apj, 671, 1160

\bibitem[{{Macci{\`o}} {et~al.}(2008){Macci{\`o}}, {Dutton}, \& {van den
  Bosch}}]{Maccio2008}
{Macci{\`o}}, A.~V., {Dutton}, A.~A., \& {van den Bosch}, F.~C. 2008, \mnras,
  391, 1940

\bibitem[{{Macci{\`o}} {et~al.}(2007){Macci{\`o}}, {Dutton}, {van den Bosch},
  {Moore}, {Potter}, \& {Stadel}}]{Maccio2007}
{Macci{\`o}}, A.~V., {Dutton}, A.~A., {van den Bosch}, F.~C., {Moore}, B.,
  {Potter}, D., \& {Stadel}, J. 2007, \mnras, 378, 55

\bibitem[{{Macci{\`o}} {et~al.}(2009){Macci{\`o}}, {Kang}, \&
  {Moore}}]{Maccio2009}
{Macci{\`o}}, A.~V., {Kang}, X., \& {Moore}, B. 2009, \apjl, 692, L109

\bibitem[{{Makino}(2004)}]{Makino2004}
{Makino}, J. 2004, \pasj, 56, 521

\bibitem[{{Merritt} {et~al.}(2006){Merritt}, {Graham}, {Moore}, {Diemand}, \&
  {Terzi{\'c}}}]{Merritt2006}
{Merritt}, D., {Graham}, A.~W., {Moore}, B., {Diemand}, J., \& {Terzi{\'c}}, B.
  2006, \aj, 132, 2685

\bibitem[{{Miyoshi} \& {Kihara}(1975)}]{Miyoshi1975}
{Miyoshi}, K., \& {Kihara}, T. 1975, \pasj, 27, 333

\bibitem[{{Moore} {et~al.}(1999{\natexlab{a}}){Moore}, {Ghigna}, {Governato},
  {Lake}, {Quinn}, {Stadel}, \& {Tozzi}}]{Moore1999a}
{Moore}, B., {Ghigna}, S., {Governato}, F., {Lake}, G., {Quinn}, T., {Stadel},
  J., \& {Tozzi}, P. 1999{\natexlab{a}}, \apjl, 524, L19

\bibitem[{{Moore} {et~al.}(1999{\natexlab{b}}){Moore}, {Quinn}, {Governato},
  {Stadel}, \& {Lake}}]{Moore1999}
{Moore}, B., {Quinn}, T., {Governato}, F., {Stadel}, J., \& {Lake}, G.
  1999{\natexlab{b}}, \mnras, 310, 1147

\bibitem[{{Mu{\~n}oz-Cuartas} {et~al.}(2010){Mu{\~n}oz-Cuartas}, {Macci{\`o}},
  {Gottl{\"o}ber}, \& {Dutton}}]{Munoz2010}
{Mu{\~n}oz-Cuartas}, J.~C., {Macci{\`o}}, A.~V., {Gottl{\"o}ber}, S., \&
  {Dutton}, A.~A. 2010, \mnras, 1685

\bibitem[{{Navarro} {et~al.}(1997){Navarro}, {Frenk}, \& {White}}]{Navarro1997}
{Navarro}, J.~F., {Frenk}, C.~S., \& {White}, S.~D.~M. 1997, \apj, 490, 493

\bibitem[{{Navarro} {et~al.}(2004){Navarro}, {Hayashi}, {Power}, {Jenkins},
  {Frenk}, {White}, {Springel}, {Stadel}, \& {Quinn}}]{Navarro2004}
{Navarro}, J.~F., {et~al.} 2004, \mnras, 349, 1039

\bibitem[{{Navarro} {et~al.}(2010){Navarro}, {Ludlow}, {Springel}, {Wang},
  {Vogelsberger}, {White}, {Jenkins}, {Frenk}, \& {Helmi}}]{Navarro2010}
{Navarro}, J.~F., {et~al.} 2010, \mnras, 402, 21

\bibitem[{{Neto} {et~al.}(2007){Neto}, {Gao}, {Bett}, {Cole}, {Navarro},
  {Frenk}, {White}, {Springel}, \& {Jenkins}}]{Neto2007}
{Neto}, A.~F., {et~al.} 2007, \mnras, 381, 1450

\bibitem[{{Okamoto} \& {Frenk}(2009)}]{Okamoto2009}
{Okamoto}, T., \& {Frenk}, C.~S. 2009, \mnras, 399, L174

\bibitem[{{Peacock}(1999)}]{Peacock1999}
{Peacock}, J.~A. 1999, {Cosmological Physics}

\bibitem[{{Portegies Zwart} {et~al.}(2010){Portegies Zwart}, {Ishiyama},
  {Groen}, {Nitadori}, {Makino}, {de Laat}, {McMillan}, {Hiraki}, {Harfst}, \&
  {Grosso}}]{Zwart2010}
{Portegies Zwart}, S., {et~al.} 2010, IEEE Computer, 43, 63

\bibitem[{{Power} \& {Knebe}(2006)}]{Power2006}
{Power}, C., \& {Knebe}, A. 2006, \mnras, 370, 691

\bibitem[{{Power} {et~al.}(2012){Power}, {Knebe}, \& {Knollmann}}]{Power2012}
{Power}, C., {Knebe}, A., \& {Knollmann}, S.~R. 2012, \mnras, 419, 1576

\bibitem[{{Power} {et~al.}(2003){Power}, {Navarro}, {Jenkins}, {Frenk},
  {White}, {Springel}, {Stadel}, \& {Quinn}}]{Power2003}
{Power}, C., {Navarro}, J.~F., {Jenkins}, A., {Frenk}, C.~S., {White},
  S.~D.~M., {Springel}, V., {Stadel}, J., \& {Quinn}, T. 2003, \mnras, 338, 14

\bibitem[{{Press} \& {Schechter}(1974)}]{Press1974}
{Press}, W.~H., \& {Schechter}, P. 1974, \apj, 187, 425

\bibitem[{{Prunet} {et~al.}(2008){Prunet}, {Pichon}, {Aubert}, {Pogosyan},
  {Teyssier}, \& {Gottloeber}}]{Prunet2009}
{Prunet}, S., {Pichon}, C., {Aubert}, D., {Pogosyan}, D., {Teyssier}, R., \&
  {Gottloeber}, S. 2008, \apjs, 178, 179

\bibitem[{{Reed} {et~al.}(2003){Reed}, {Gardner}, {Quinn}, {Stadel}, {Fardal},
  {Lake}, \& {Governato}}]{Reed2003}
{Reed}, D., {Gardner}, J., {Quinn}, T., {Stadel}, J., {Fardal}, M., {Lake}, G.,
  \& {Governato}, F. 2003, \mnras, 346, 565

\bibitem[{{Reed} {et~al.}(2005){Reed}, {Governato}, {Verde}, {Gardner},
  {Quinn}, {Stadel}, {Merritt}, \& {Lake}}]{Reed2005b}
{Reed}, D., {Governato}, F., {Verde}, L., {Gardner}, J., {Quinn}, T., {Stadel},
  J., {Merritt}, D., \& {Lake}, G. 2005, \mnras, 357, 82

\bibitem[{{Reed} {et~al.}(2007){Reed}, {Bower}, {Frenk}, {Jenkins}, \&
  {Theuns}}]{Reed2007}
{Reed}, D.~S., {Bower}, R., {Frenk}, C.~S., {Jenkins}, A., \& {Theuns}, T.
  2007, \mnras, 374, 2

\bibitem[{{Reed} {et~al.}(2011){Reed}, {Koushiappas}, \& {Gao}}]{Reed2011}
{Reed}, D.~S., {Koushiappas}, S.~M., \& {Gao}, L. 2011, \mnras, 415, 3177

\bibitem[{{Sheth} \& {Tormen}(1999)}]{Sheth1999}
{Sheth}, R.~K., \& {Tormen}, G. 1999, \mnras, 308, 119

\bibitem[{{Springel} {et~al.}(2005){Springel}, {White}, {Jenkins}, {Frenk},
  {Yoshida}, {Gao}, {Navarro}, {Thacker}, {Croton}, {Helly}, {Peacock}, {Cole},
  {Thomas}, {Couchman}, {Evrard}, {Colberg}, \& {Pearce}}]{Springel2005}
{Springel}, V., {et~al.} 2005, \nat, 435, 629

\bibitem[{{Springel} {et~al.}(2008){Springel}, {Wang}, {Vogelsberger},
  {Ludlow}, {Jenkins}, {Helmi}, {Navarro}, {Frenk}, \& {White}}]{Springel2008}
{Springel}, V., {et~al.} 2008, \mnras, 391, 1685

\bibitem[{{Stadel} {et~al.}(2009){Stadel}, {Potter}, {Moore}, {Diemand},
  {Madau}, {Zemp}, {Kuhlen}, \& {Quilis}}]{Stadel2009}
{Stadel}, J., {Potter}, D., {Moore}, B., {Diemand}, J., {Madau}, P., {Zemp},
  M., {Kuhlen}, M., \& {Quilis}, V. 2009, \mnras, 398, L21

\bibitem[{{Strigari} {et~al.}(2008){Strigari}, {Bullock}, {Kaplinghat},
  {Simon}, {Geha}, {Willman}, \& {Walker}}]{Strigari2008}
{Strigari}, L.~E., {Bullock}, J.~S., {Kaplinghat}, M., {Simon}, J.~D., {Geha},
  M., {Willman}, B., \& {Walker}, M.~G. 2008, \nat, 454, 1096

\bibitem[{{Taylor} \& {Navarro}(2001)}]{Taylor2001}
{Taylor}, J.~E., \& {Navarro}, J.~F. 2001, \apj, 563, 483

\bibitem[{{Teyssier} {et~al.}(2009){Teyssier}, {Pires}, {Prunet}, {Aubert},
  {Pichon}, {Amara}, {Benabed}, {Colombi}, {Refregier}, \&
  {Starck}}]{Teyssier2009}
{Teyssier}, R., {et~al.} 2009, \aap, 497, 335

\bibitem[{{Tinker} {et~al.}(2008){Tinker}, {Kravtsov}, {Klypin}, {Abazajian},
  {Warren}, {Yepes}, {Gottl{\"o}ber}, \& {Holz}}]{Tinker2008}
{Tinker}, J., {Kravtsov}, A.~V., {Klypin}, A., {Abazajian}, K., {Warren}, M.,
  {Yepes}, G., {Gottl{\"o}ber}, S., \& {Holz}, D.~E. 2008, \apj, 688, 709

\bibitem[{{van den Bosch} {et~al.}(2005){van den Bosch}, {Tormen}, \&
  {Giocoli}}]{Bosch2005}
{van den Bosch}, F.~C., {Tormen}, G., \& {Giocoli}, C. 2005, \mnras, 359, 1029

\bibitem[{{Wambsganss} {et~al.}(2004){Wambsganss}, {Bode}, \&
  {Ostriker}}]{Wambsganss2004}
{Wambsganss}, J., {Bode}, P., \& {Ostriker}, J.~P. 2004, \apjl, 606, L93

\bibitem[{{Wang} {et~al.}(2011){Wang}, {Mo}, {Jing}, {Yang}, \&
  {Wang}}]{Wang2011}
{Wang}, H., {Mo}, H.~J., {Jing}, Y.~P., {Yang}, X., \& {Wang}, Y. 2011, \mnras,
  413, 1973

\bibitem[{{Warren} {et~al.}(2006){Warren}, {Abazajian}, {Holz}, \&
  {Teodoro}}]{Warren2006}
{Warren}, M.~S., {Abazajian}, K., {Holz}, D.~E., \& {Teodoro}, L. 2006, \apj,
  646, 881

\bibitem[{{Wechsler} {et~al.}(2002){Wechsler}, {Bullock}, {Primack},
  {Kravtsov}, \& {Dekel}}]{Wechsler2002}
{Wechsler}, R.~H., {Bullock}, J.~S., {Primack}, J.~R., {Kravtsov}, A.~V., \&
  {Dekel}, A. 2002, \apj, 568, 52

\bibitem[{{White} {et~al.}(2010){White}, {Cohn}, \& {Smit}}]{White2010}
{White}, M., {Cohn}, J.~D., \& {Smit}, R. 2010, \mnras, 408, 1818

\bibitem[{{White} \& {Rees}(1978)}]{White1978}
{White}, S.~D.~M., \& {Rees}, M.~J. 1978, \mnras, 183, 341

\bibitem[{{Yahagi} {et~al.}(2004){Yahagi}, {Nagashima}, \&
  {Yoshii}}]{Yahagi2004}
{Yahagi}, H., {Nagashima}, M., \& {Yoshii}, Y. 2004, \apj, 605, 709

\bibitem[{{Yoshikawa} \& {Fukushige}(2005)}]{Yoshikawa2005}
{Yoshikawa}, K., \& {Fukushige}, T. 2005, \pasj, 57, 849

\bibitem[{{Zentner} {et~al.}(2005){Zentner}, {Berlind}, {Bullock}, {Kravtsov},
  \& {Wechsler}}]{Zentner2005}
{Zentner}, A.~R., {Berlind}, A.~A., {Bullock}, J.~S., {Kravtsov}, A.~V., \&
  {Wechsler}, R.~H. 2005, \apj, 624, 505

\bibitem[{{Zhao} {et~al.}(2003){Zhao}, {Jing}, {Mo}, \&
  {B{\"o}rner}}]{Zhao2003}
{Zhao}, D.~H., {Jing}, Y.~P., {Mo}, H.~J., \& {B{\"o}rner}, G. 2003, \apjl,
  597, L9

\bibitem[{{Zhao} {et~al.}(2009){Zhao}, {Jing}, {Mo}, \&
  {B{\"o}rner}}]{Zhao2009}
{Zhao}, D.~H., {Jing}, Y.~P., {Mo}, H.~J., \& {B{\"o}rner}, G. 2009, \apj, 707,
  354

\end{thebibliography}
\end{document}